\documentstyle[preprint,eqsecnum,epsf,prb,aps]{revtex}
\begin{document}
\preprint{IUCM-95046}
\draft
\title{Scaling theory of the Kondo screening cloud}
\author{Erik S.\ S\o rensen}
\address{Department~of~Physics,
Indiana~University, Bloomington, IN~47405}
\author{Ian~Affleck}
\address{Department of Physics and Canadian Institute
for Advanced Research}
\address{
University of British Columbia, Vancouver, BC, V6T 1Z1, Canada}
\date{\today}
\maketitle
\begin{abstract}
A scaling form for the local susceptibility, derived
from renormalization group arguments, is proposed. The
scale over which
the uniform part of this scaling form varies can be viewed
as a definition of the Kondo ``screening cloud" $\sim \xi_K$.
The proposed scaling
form interpolates between Ruderman-Kittel-Kasuya-Yosida
(RKKY) results in the high temperature
limit, $T\gg T_K$, and Fermi liquid results in the low
temperature, long-distance limit, $T\ll T_K$, $r\gg \xi_K$.
The predicted form of the Knight shift is longer range
at low temperatures where the screening cloud has formed,
than at high temperatures where it has not.
Using weak and strong coupling perturbation theory combined
with large scale density matrix renormalization group (DMRG) results we
study the validity of the finite size version of the scaling form at $T=0$.
We explicitly extract a length scale proportional to the
Kondo length scale, $\xi_K$.
The numerical results are in good agreement with the proposed scaling form
and confirm the existence of the Kondo screening cloud.
\end{abstract}
\pacs{75.20.Hr, 75.10.Lp, 75.40.Mg}
%============================================================================
% BODY OF PAPER

\section{Introduction}
The Kondo effect is probably one
of the most well-studied phenomena in condensed matter physics.  The
highly successful theoretical approaches  include Wilson's numerical
renormalization group (NRG) method~\cite{wilson}, simpler and more physical
renormalization group (RG) approaches of Anderson et al.~\cite{pwa}
and Nozi\`eres~\cite{nozieres} and exact Bethe ansatz results~\cite{bethe}.
Thus it is perhaps surprising that a
fundamental aspect of this problem remains mired in controversy.

{}From the RG viewpoint, the Kondo effect is associated with very large
distance scales, $\xi_K \approx ae^{1/\rho J}$, where $a$ is the lattice
spacing, $\rho$ the electronic density of states at the Fermi surface
and $J$ the Kondo coupling.  This scale is essentially $\xi_K \approx
v_F/T_K\approx aE_F/T_K$, where $v_F$ is the Fermi velocity, $E_F$ the
{}Fermi energy and $T_K$, the Kondo temperature, is the energy scale
associated with the Kondo effect.  Since Kondo temperatures are normally
of order 10's of degrees, this scale is normally thousands of lattice
spacings (ie. microns).  A heuristic description of the RG results on
the Kondo problem says that a cloud of electrons of this order of
magnitude surrounds the impurity spin, forming a singlet with it.  The
remaining low-energy electronic excitations outside the screening cloud
do not ``feel'' the impurity spin.  Rather the screened complex acts
like a potential-scatterer for these electrons, with a unitary limit
phase shift of $\pi /2$ right at the Fermi energy.  The largeness of
this Kondo length scale, $\xi_K$ , in experimental systems in which the
Kondo effect is apparently observed is rather disconcerting.  Even a
very dilute system with 1 part per million of impurities has a typical
inter-impurity separation of about 100 lattice spacings, much smaller
than $\xi_K$.  Thus each impurity has many other impurities inside its
screening cloud and it is surprising that the single-impurity Kondo
effect is observed at all.  Nonetheless, the impurity resistivity,
susceptibility, etc. are observed to be linear in impurity concentration
and these quantities seem to fit theoretical expectations.

Comparatively little of the theoretical work on the Kondo effect has
focussed on spatial correlations.  These seem to be difficult to obtain using
Wilson's method and impossible from the Bethe ansatz.  Perturbative
calculations have been
performed~\cite{Bressmann,Fullenbaum,keiter,Gan}
as have calculations using the ``Nagoaka equations'' a type of
mean field theory~\cite{mhartman}.  Renormalization group approaches have
been developed by Chen et al.~\cite{chen} and Gan~\cite{Gan}.
However, Chen et al.~\cite{chen} only
consider short-range correlations, with $r\ll \xi_K$ and do not address the
issue of the size of the screening cloud.  We incorporate the perturbative
results into our discussion of the renormalization
group and scaling, but come
to rather different conclusions than Gan~\cite{Gan}
about the scaling variables and
the size of the screening cloud.  Related theoretical work has addressed the
screening cloud in the Anderson model using various approximate
methods~\cite{gubernatis,sokcevic,ishii}. Some of the previous results
have been reviewed in Ref.~\onlinecite{kondo}, and \onlinecite{hewson}.

Experiments which have attempted to
look for this large screening cloud have obtained
mixed results~\cite{slichter,nmr}. In particular, the NMR experiments of
Boyce and Slichter~\cite{slichter}
were interpreted to indicate the absence of this cloud.  This has led to
some theoretical discussion about the circumstances under which this
cloud can be observed and even to some doubts about its existence.

The purpose of this paper is to examine in more detail, the behavior of
the Knight shift (ie. the  electronic spin polarization by an applied
field) in the vicinity of a magnetic impurity.  In the next section we
make a scaling hypothesis about this quantity based on standard RG
arguments and assuming the existence of a large screening cloud. We
point out the rather unintuitive result that the Knight shift is
actually longer range at low temperatures, where the screening cloud has
formed, than at high temperatures where it has not.   We argue that the
NMR experiments are not necessarily in contradiction with our scaling
form.  The basic problem with the experiments, according to this view,
is that they only probed very  short distances, $< 3a$, whereas $\xi_K$
is presumably thousands of times larger than $a$.

We then test our scaling hypothesis numerically.  This is done using a
one-dimensional tight-binding model.  We don't expect that the reduced
dimensionality is important since the Kondo problem is intrinsically
one-dimensional anyway.  Taking a spherically symmetric dispersion
relation and a $\delta$-function Kondo interaction, we may decompose the
electronic degrees of freedom into spherical harmonics.  Only the s-wave
interacts, and this corresponds to a one-dimensional problem.  The
numerical method we use basically restricts us to $T=0$ and a finite
length, $L\leq 40-50$.  As will be seen, this finite length plays
essentially the role of an inverse temperature in our scaling arguments.
Both the reduced dimensionality and the finite length may be directly
relevant to more recent experiments~\cite{lowD}
which have attempted to find the
screening cloud using small samples with lengths of order
$\xi_K$ or smaller. Recent theoretical work has
also addressed these issues~\cite{lowDtheo}.

In Section~\ref{sec:rg} we present the scaling form for
the local susceptibility. Renormalization group arguments are
given bridging a high temperature RKKY form with the low temperature
long distance form in a single scaling expression. Sections~\ref{sec:tb} and
\ref{sec:num} briefly discuss
the form of the Hamiltonian we use
in our numerical work and some details of the
numerical method. Weak and strong coupling perturbation results are presented
in section~\ref{sec:weak} and \ref{sec:strong}
along with DMRG results. Finally in Section~\ref{sec:cross}
the cross over region is studied and the scaling form is tested.

\section{Renormalization Group Arguments}\label{sec:rg}

We
consider the standard Kondo model:
\begin{equation}
H =
\sum_{\bf k}\epsilon_k \psi^{\dagger \alpha}_{\bf  k}
\psi^{\vphantom{\dagger \alpha}}_{{\bf k} \alpha} +
J{\bf S}_{{\rm imp}}\cdot \sum_{\bf  k,\bf  k'}\psi^{\dagger \alpha}_{\bf  k}
\frac{\bbox{\sigma}^{\beta}_{\alpha}}{2}
\psi^{\vphantom{\dagger \alpha}}_{{\bf  k'} \beta}.
\label{eq:hkk}
\end{equation}
Here $\psi_{\bf  k \alpha}$
is the annihilation operator for conduction electrons of momentum ${\bf
k}$, spin $\alpha$.  In the following we suppress
spin indices which are implicitly summed over.  ${\bf S}_{{\rm imp}}$ is
the impurity spin operator of magnitude $s=1/2$.
The $\sigma^a$'s are Pauli matrices, and we set $\hbar =1$. The total
spin operator is:
\begin{equation}
{\bf S}_{{\rm tot}} = {\bf
S}_{{\rm imp}} + \sum_{\bf  k} \psi^\dagger_{\bf  k}\frac{\bbox{\sigma}}{2}
\psi^{\vphantom{\dagger}}_{\bf  k}.
\label{eq:stot}
\end{equation}
(We assume, for simplicity, equal g-factors
for the impurity spin and conduction electrons.) The Knight shift is
proportional to the local susceptibility:
\begin{equation}
\chi (r, T)
\equiv (1/T)<\psi^\dagger ({\bf r}) {\sigma^z\over 2}\psi ({\bf r})
S^z_{{\rm tot}}>.
\label{eq:locsus}
\end{equation}
This consists of a bulk part, the usual Pauli susceptibility, $\rho /2$,
where $\rho$ is the density of states per spin, together with a local part
arising from the impurity.  The total change in the susceptibility due to
the impurity, usually called the impurity susceptibility,
$\chi_{{\rm imp}}$, is: \begin{equation}
\chi_{{\rm imp}} = \int d^3r [\chi (r)-\rho /2] +
(1/T)<S^z_{{\rm imp}}S^z_{{\rm
tot}}>.\label{chi_imp}\end{equation}

 Lowest order perturbation theory gives the RKKY
result, which becomes, at $rk_F\gg 1$:  \begin{equation}
\chi (r,T)
= \frac{\rho}{2} + {\lambda \over 16r^2v_F\sinh {2\pi rT\over  v_F}}\cos
2k_Fr.
\label{RKKY}
\end{equation}
Here
$\lambda$ is the  dimensionless coupling constant,
\begin{equation}
\lambda \equiv \rho J.
\end{equation}
In the limit, $r\ll v_F/T$, (but still $r\gg 1/k_F$) this gives
the well-known RKKY expression:
\begin{equation}
\chi -\frac{\rho}{2}
\to {\lambda \over 32\pi r^3T}\cos 2k_Fr.
\label{RKKY0}
\end{equation}

A crucial feature of the Kondo problem is that, for antiferromagnetic
coupling, the Kondo coupling increases under renormalization as the
energy scale is reduced.  The lowest order renormalization group
equation:
\begin{equation}
d\lambda /d \ln \Lambda =
-\lambda^2,
\end{equation}
is obtained.  Here $\Lambda$ is the momentum
space  cut-off, or effective band-width.  This gives the effective
coupling at momentum scale $\Lambda$:
\begin{equation}
\lambda_{{\rm eff}}(\Lambda ) = {\lambda \over 1 -\lambda \ln
(\Lambda_0/\Lambda )}.
\label{lambda_eff}
\end{equation}
Here
$\Lambda_0\approx 1/a$ is the bare cut-off and $\lambda$ is the bare
coupling (defined at that scale).  The Kondo length scale is defined
from the momentum scale at which the effective Kondo coupling constant
diverges:
\begin{equation}
\xi_K=v_F/T_K \approx \Lambda_0^{-1}
e^{1/\lambda}.
\label{xi_K}
\end{equation}
A finite temperature acts as an infrared cut-off on perturbation theory
so that Eq. (\ref{lambda_eff}) with $\Lambda$ replaced by $T/v_F$ can be
used to define a temperature-dependent effective coupling.

Eq. (\ref{RKKY}) is only  valid at high temperatures and weak Kondo
coupling.  As the temperature is lowered, the effective Kondo coupling
increases so higher order terms become important.  The corrections of
$O(\lambda^2)$ has been calculated.
{}From Eq.~(\ref{eq:stot}), we see that $\chi
(r,T)$ is a sum of two terms:
\begin{eqnarray}
\chi (r) &=& \int_0^\beta d\tau <\psi^\dagger ({\bf r},0)
{\sigma^z\over 2}\psi ({\bf r})S^z_{{\rm imp}}(\tau )> \nonumber\\
& &+
\int_0^\beta d\tau <\psi^\dagger
({\bf r}) {\sigma^z\over 2}\psi ({\bf r},0)\int d^3r' \psi^\dagger
({\bf r}',\tau ) {\sigma^z\over 2}\psi ({\bf r}',\tau )>\nonumber \\&\equiv &
\chi_{de}(r) + \chi_e(r).
\end{eqnarray}
We have adopted the notation of Ref. (\onlinecite{Gan}).
Note that the sum of these correlation functions is
independent of $\tau$ since the total spin is conserved, so that the
$\tau$-integral simply gives a factor of $\beta$.  However, the individual
correlation functions depend non-trivially on $\tau$.  We could equally
well add the equal time correlation functions and multiply by $\beta$, but
we choose the above representation because both terms have been evaluated
explicitly in the literature, in a convenient form.  The RKKY term of
$O(\lambda )$ in eq. (\ref{RKKY}), comes entirely from $\chi_{de}$.
In the asymptotic region, $rk_F\gg 1$, $E_F/T\gg 1$,
with $r\ll v_F/T$, including the correction of
$O(\lambda^2)$:
\begin{equation}
\chi (r) -\frac{\rho}{2} =
{\cos 2k_Fr\over 32\pi r^3T} \{\lambda +\lambda^2
[\ln (k_Fr)+{\rm constant}]\}.\label{chiren}
\end{equation}
The logarithmic
term comes entirely from $\chi_{de}$.  It was
first calculated in Ref. (\onlinecite{Fullenbaum}) and
(\onlinecite{Bressmann}).   $\chi_e$ contributes only to the constant in
Eq. (\ref{chiren}).  [See Eq. (B2) of Ref. (\onlinecite{Gan}).]

Note, from Eq. (\ref{lambda_eff}), that the quantity in brackets in
Eq.~(\ref{chiren}) may be written $\lambda_{\rm eff}(r) + \lambda_{\rm
eff}(r)^2+ {\rm constant}$, to $O(\lambda^2)$.  This expression exhibits an
infrared divergence at large $r$.  That is, for sufficiently large $r$,
$r>\xi_K$, the $O(\lambda^2)$ term exceeds the $O(\lambda )$ term.  Note
however, that this correction term is at least finite as $T\to 0$.  ie. it
is $\lambda_{\rm eff}(r)$ that appears, not
$\lambda_{\rm eff}(T)$.  Thus, at least to $O(\lambda^2)$, a
finite $r$ is acting like a cut-off on the infrared divergences of
perturbation theory.  It is an important question whether or not this
persists to higher orders in perturbation theory.  ie., is
perturbation theory valid for $r\ll \xi_K$ even for $T\ll T_K$, with the
actual expansion parameter being $\lambda_{\rm eff}(r)$?  Based
on an examination of higher order terms Gan\cite{Gan} has argued this
not to be the case.  He claims that higher order terms diverge as $T
\to 0$ for non-zero $r$ and that it is therefore necessary to have
$T\gg T_K$ for perturbation theory to be valid. [This point will be
examined in detail in Ref. (\onlinecite{Barzykin}).] In this case, it is
probably more useful to rewrite Eq.~(\ref{chiren}) in terms of
$\lambda_{\rm eff}(T)$. To $O(\lambda^2)$:  \begin{equation}
 \chi (r) -\frac{\rho}{2} =
{\cos 2k_Fr\over 32\pi r^3T} \{\lambda_{eff}(T)
+\lambda_{eff}(T)^2[ \ln
(rT/v_F)+{\rm constant}]\},
\label{chiren2}
\end{equation}
using
$E_F/k_F\approx v_F$.

At very low
temperatures and large distances, $T\ll T_K$, $r\gg \xi_K$, we expect $\chi
(r)$ to be determined by the zero-energy fixed point.
Within the local Fermi liquid theory~\cite{nozieres}
of this fixed point we can then
estimate $\chi(r)$.
The zero-energy fixed point corresponds
to a screened impurity which just acts as a potential scatterer for the
low energy electronic degrees of freedom, with a phase shift of $\pi /2$
at the Fermi energy.  The local susceptibility of a potential scatterer
follows directly from the formula for Friedel oscillations in the
electron density, $n(r)$, with an s-wave scatterer and a $\pi /2$ phase
shift. For $k_Fr\gg 1$:
\begin{equation}
n (r) = n_0 - {1\over
2\pi^2r^3}\cos [2k_Fr+ \pi /2 ].
\label{eq:nr}
\end{equation}
Noting that a magnetic
field, $H$, simply shifts the chemical potential by $\pm g\mu_BH/2$ for
spin down or spin up electrons, we obtain:
\begin{equation}
\chi
(r,T)={1\over 4v_F}{dn \over dk_F} = {\rho \over 2} + {1\over
4\pi^2v_Fr^2}\cos (2k_Fr).
\label{chilt}
\end{equation}
Note that $\chi
(r,T)$ is longer-range at low T after the screening cloud has formed,
$\chi \propto 1/r^2$, than at higher T before it has formed, $\chi
\propto 1/r^3$, (Eq.~(\ref{RKKY})). An analogous result occurs in
spin chain systems~\cite{sebastian}.

Corrections to Eq.~(\ref{chilt}) can be derived by doing perturbation
theory in the leading irrelevant operator.
Part of the leading
correction can be obtained by considering a field-dependence of the
phase shift:\cite{nozieres}
\begin{equation}
\delta^\sigma = \pi /2 + \sigma hc/T_K,
\end{equation}
where $c$ is a dimensionless constant of $O(1)$.  Generalizing the
{}Friedel oscillation formula of Eq.~(\ref{eq:nr}), the local density of
spin-$\sigma$ electrons becomes:
\begin{equation}
n_\sigma (r) = {n_0\over 2} - {1\over 4\pi r^3}\cos
[2k_F^\sigma (h)r+\delta^\sigma (h) ].
\label{sub}
\end{equation}
Upon
differentiating with respect to $h$ to obtain the local
susceptibility, we now obtain an additional term:
\begin{equation}
\delta \chi = {c\over 4\pi^2r^3T_K}\cos
(2k_Fr).
\label{eq:dchi}
\end{equation}
Note that this term drops off more rapidly
with $r$ than the term in Eq. (\ref{chilt}) obtained from differentiating
$k_F^\sigma(h)$ and is smaller for $r\gg \xi_K$.
A very similar Fermi
liquid calculation of $\chi (r)$, in the Anderson model, was performed
in Ref. \onlinecite{ishii}.   However, this calculation effectively
ignored the field dependence of $k_F^\sigma$, and hence only obtained
the subdominant term of Eq. (\ref{eq:dchi}), not the leading term of
Eq.~(\ref{chilt}).
Explicitly, in Eq. (2.10) of Ref. \onlinecite{ishii}, the free
electron Green's function, $F_r(i\omega_l)$ must be evaluated in a
finite magnetic field.  Taking this into account, we obtain our
expression, Eq.~(\ref{chilt}).

According to Fermi liquid theory,
the impurity susceptibility, $\chi_{{\rm imp}}$, defined in Eq.
(\ref{chi_imp}) is $O(1/T_K)$.  This appears to arise from a
short-range part of $\chi (r)$ which does not oscillate at wave-vector
$2k_F$.  However, since the impurity has been ``integrated out'' to obtain
the Fermi liquid theory, it is difficult to ascertain how much of
$\chi_{{\rm imp}}$ comes from  $\chi (r)$ and how much comes from
the impurity self-correlation function, in Eq. (\ref{chi_imp}). It is also
difficult to tell whether the contribution, if any, from $\chi (r)$
has a range of $O(\xi_K)$ or only of $O(1/k_F)$, because the cut-off has
been reduced to $O(1/\xi_K)$ to obtain the Fermi liquid theory.

We now wish to formulate a scaling hypothesis for $\chi (r,T)$ which we
expect to be valid at arbitrary $r$ and $T$ in the scaling region,
$r\gg a$, $T\ll E_F$.
{}For this purpose it is very convenient to use the
relativistic one-dimensional formulation of the Kondo problem.  [Ref.
{}~\onlinecite{ludwig}]  The mapping to one-dimension is exact for pure s-wave
scattering.  The use of a reduced bandwidth and linear dispersion
relation which leads to the relativistic model is expected to be
valid in the scaling region.  The three dimensional electron field is
expanded in spherical harmonics and then the s-wave part is written
in terms of left and right moving components (ie. incoming and
outgoing):
\begin{equation}
\psi ({\bf x}) = {1\over 2\sqrt{2}\pi
r}\left[e^{-ik_Fr}\psi_L(r)-e^{ik_Fr}\psi_R(r)\right]
+ {\rm higher\  harmonics}.
\end{equation}
The left and right-moving fields, defined
on $r>0$ obey the boundary condition:
\begin{equation}
\psi_L(0)=\psi_R(0).
\end{equation}
We may flip the
right-moving field to the negative axis, so that we work with
left-movers only defined on the entire real axis:
\begin{equation}
\psi_L(-x)\equiv \psi_R(x).
\end{equation}
The
one-dimensional Hamiltonian can be written:
\begin{equation}
H=v_F\int_{-\infty}^\infty
dr\psi^\dagger_L(r)(id/dr)\psi_L(r) + v_F\lambda
\psi^\dagger_L(0){\bbox{\sigma} \over 2}\psi_L(0)\cdot
{\bf S}_{\rm imp}.
\end{equation}
$\chi
-\rho /2$ can be expanded in spherical harmonics; only the s-wave
harmonic is non-zero. This can be written in terms of one-dimensional
uniform and $2k_F$ susceptibilities:
\begin{equation}
\chi -\rho /2 = {1\over
8\pi^2r^2}[\chi_{\rm un}+(e^{2ik_Fr}\chi_{2k_F}+c.c.)],
\label{scaling1}
\end{equation}
where $c.c.$ denotes complex conjugate and:
\begin{eqnarray}
\chi_{un}(r,T) &\equiv&
(1/T)<[\psi^\dagger_L(r){\sigma^z\over
2}\psi_L(r)+\psi^\dagger_L(-r){\sigma^z\over
2}\psi_L(-r)]S^z_{T}>\nonumber \\
\chi_{2k_F}(r,T) &\equiv&
(1/T)<\psi^\dagger_L(r){\sigma^z\over
2}\psi_L(-r)S^z_{T}>.\end{eqnarray}
Here ${\bf S}_{T}$ is the total spin in the one-dimensional
theory:
\begin{equation}
{\bf S}_{T} \equiv
{\bf S}_{{\rm imp}}+{1\over 2\pi}\int_{-\infty}^\infty
dr\psi^\dagger_L(r){ \bbox{\sigma} \over 2}\psi_L(r).
\end{equation}

$\chi_{2k_F}$ can be shown to be real using particle-hole symmetry.
This follows since under particle-hole symmetry:
\begin{equation}
\psi_L(r)\to \sigma^y\psi_L^\dagger (r),
\end{equation}
and hence:
\begin{eqnarray}
{\bf S}_{T}&\to& {\bf S}_{T}
\nonumber \\
\psi^\dagger_L(r)\bbox{\sigma} \psi_L(-r)&\to&
\psi^\dagger_L(-r)\bbox{\sigma} \psi_L(r)
=\left[ \psi^\dagger_L(r)\bbox{\sigma} \psi_L(-r)\right]^\dagger .
\end{eqnarray}
When particle-hole symmetry is broken, as it is for a realistic
Hamiltonian, we expect $\chi_{2k_F}$ to have a phase, $\theta$, which is
non-zero but constant in the scaling region.  This can be seen from
spin-charge separation in the one-dimensional formulation of the Kondo
problem.  The Kondo interaction, which produces the non-trivial scaling
behavior, occurs entirely in the spin sector.  In the absence of
particle-hole
symmetry there is a marginal potential scattering term, $-\theta
\psi^\dagger_L(0)\psi_L(0)$, which is a pure charge operator.  Upon
bosonizing, this is linear in the charge boson and hence doesn't
renormalize (is exactly marginal). We can write
$\chi_{2k_F}$ as a product of spin and charge correlation functions.
The charge correlation function just contributes a constant factor
$e^{i\theta}$ to $\chi_{k_F}$.

In the lowest two orders of perturbation theory, discussed above, the
function $\chi_{\rm un}$ vanishes at $r\gg 1/k_F$.  In fact, it is possible
to prove that this happens to all orders in perturbation
theory.\cite{Barzykin}
This is also consistent with Fermi liquid theory, given
the uncertainties in that theory, discussed above, about the origin of
$\chi_{{\rm imp}}$.

We expect the  one-dimensional local susceptibility
to obey scaling in the following sense.  After extracting a factor of
$1/v_F$, $\chi_{2k_F}$ could, in principle depend on three dimensionless
variables, which can we taken to be, $rT/v_F$, $\lambda$ and $D/T$.
Here $D$
is the effective bandwidth in the one-dimensional theory, a quantity of
$O(E_F)$.  The scaling hypothesis asserts that the bare
coupling constant, $\lambda$ and $D/T$ do not appear independently
but only in the combination making up the renormalized coupling constant,
$\lambda_{\rm eff}(T)$. The dependence on $\lambda_{\rm eff}(T)$
may be exchanged for a dependence on $T/T_K$.
To see that these two quantities are related, note, from
Eqs. (\ref{lambda_eff}), and (\ref{xi_K}), valid at weak
$\lambda_{{\rm eff}}$,
\begin{equation}
\exp
\left[{1\over\lambda_{{\rm eff}}(T)}\right] = T/T_K.
\end{equation}
In the intermediate to strong coupling region, the value of
$\lambda_{{\rm eff}}$ is non-universal, ie. ambiguous.  In this regime
it is better to use $T/T_K$ as a measure of the dimensionless effective
coupling.  This is in accord with the idea that $1/T_K$ is the coupling
constant for the leading irrelevant operator at the low-temperature
fixed point.  As usual, we multiply by the effective cut-off, $T$, to
form the dimensionless coupling constant. Thus we write the scaling
hypothesis as:
\begin{equation}
\chi_{2k_F} = {1\over
v_F}f(rT/v_F,T/T_K),
\label{scaling}
\end{equation}  where $f$ is a real universal scaling function.
Using Eq. (\ref{scaling1}) the equivalent statement for the
three-dimensional susceptibility is:
\begin{equation}
\chi -\rho /2 = {1\over 8\pi^2v_Fr^2}\cos (2k_Fr)f(rT/v_F,
T/T_K).\ \ \ (r\gg 1/k_F; T,T_K\ll E_F) \label{scaling3}\end{equation}

As we will see
below, this is consistent with what is known about the behavior at
$T\gg T_K$ from perturbation theory and $T\ll T_K$ and $r\gg \xi_K$ from the
local Fermi liquid description of the critical point. Note that this
scaling hypothesis {\it does} imply the existence of the large screening
cloud, since if $T\leq T_K$, the length scale over which $\chi (r)$ varies
is at least $\xi_K$ (apart from the $2k_F$ oscillations and the short-range
part).
In fact, this scaling hypothesis is perhaps the best definition of what
it means to have a screening cloud. Note that this scaling form does not
include any anomalous dimension.  We expect this to be absent since
${\bf S}_{\rm tot}$ is conserved.  The more general case, with unequal
gyromagnetic ratios for electrons and impurity, involves a non-conserved
operator.  This  will be discussed in Ref. (\onlinecite{Barzykin}).

The known results from perturbation theory and Fermi liquid theory, Eq.
(\ref{RKKY}), (\ref{chiren2}), (\ref{chilt}) and (\ref{eq:dchi}),
are all consistent with this
scaling hypothesis and imply certain limiting forms for the scaling
function.  Eq. (\ref{RKKY}) implies that
\begin{equation}
f(x,y) \to {\pi^2 \over 2\sinh (2\pi x)\ln y},
\label{faslT}
\end{equation} for $y\ll 1$.
Eq. (\ref{chiren2}) gives a higher
order correction in $1/\ln y$ to f, when $x$ is also small. Eq.
(\ref{chilt}) and (\ref{eq:dchi}) imply that, for $y\ll 1$ and $x/y\gg 1$,
\begin{equation}
f(x,y) \to 2 + {\rm constant}\cdot {y\over x}.
\end{equation}

The function $f$ in the regime $y\leq 1$, $x/y =
r/\xi_K\leq 1$ is of special interest.  It describes the interior of
the screening cloud at low T.
One might naively suppose that a small
$r\ll \xi_K$ would also cut off the renormalization of the effective
coupling so that deep inside the screening cloud we recover
weak-coupling behavior (for weak bare coupling) even at low $T$.   As
mentioned above, Gan\cite{Gan} has argued, based on higher order
perturbative calculations, that this is not the case.  It follows that
the scaling function would be non-trivial in this region.

Now let us consider the experiments of Boyce and Slichter on Fe doped
Cu. They measured what they interpreted as the Knight shift from 5
different shells of Cu atoms at distances up to 5th nearest neighbor.
We note that for the Cu fcc lattice, assuming the Fe impurities occupy Cu
lattice sites, the fifth nearest neighbor is at a distance of
$\sqrt{6}a\approx 2.4a$ where $a$ is the nearest neighbor separation.
The measurements were taken  from T=300K down to well below what is
believed to be the Kondo temperature of 29K.  They found the factorized
form:
\begin{equation}
\chi (r,T) = {f(r)\over T+T_K},
\end{equation}
for
some rapidly varying function $f(r)$ (which, in fact, changes sign over
the small range of r considered).  Note that all measurements are taken
in the regime $rT/v_F\ll 1$, $rT_K/v_F\ll 1$.  In fact the values of $r$ are
so small that it is unclear whether the scaling form of Eqs.
(\ref{scaling3}) holds at all.  In particular, the
short-range part of $\chi$ may be contributing.  If we assume $r$ is large
enough that this can be ignored, and the scaling form holds, then we may
consider the short distance limit $r\ll v_F/T, v_F/T_K$ of the scaling
function.    According to Gan\cite{Gan} the
behavior of $f(x,y)$ is non-trivial in the lower temperature range of the
experiment, $y\ll 1$, $x/y\ll 1$.  ie., we do not know the behavior
of the scaling function at low $T$ deep inside the screening cloud.
It is possible that $f(x,y)$ exhibits an approximately factorized form
for $x/y\ll 1$ and all $y$:
\begin{equation}
f(x,y) \approx {f(x/y)\over
y+1} ??
\end{equation}  This behavior would explain the experimental results.
  Note that such
factorization could not also occur at large r, $r\gg \xi_K$ if our
assumed scaling and asymptotic behaviors are correct.  In this region,
the $1/r^3$ behavior at $T\gg T_K$ crosses over to
$1/r^2$ at $T\ll T_K$. More experiments at larger $r$ could clarify the
situation.  Experiments in the region $1/k_F\ll r\ll \xi_K$ may be feasible
.  These would probe the short-distance part of the scaling function.
A full study of the scaling function would require going out to values
of $r\geq \xi_K$. One point to
bear in mind is that, assuming the existence of a large screening cloud,
the average impurity separation is much less than $\xi_K$ so
inter-impurity interactions may be playing a large role.  This may make
experimental observation of the screening cloud very difficult, at least
until a better understanding of the effect of inter-impurity
interactions is obtained.

\section{Tight Binding Model}\label{sec:tb}
In order to apply the
density matrix renormalization group
(DMRG) method we rewrite Eq.~(\ref{eq:hkk}) in real space.
This is the standard $s-d$ Kondo model. The model is described by
a tight binding Hamiltonian coupled to a $s=1/2$ impurity
spin, ${\bf S}_{\rm imp}$.
\begin{equation}
H= -t\sum_{i=1}^{L-1}\left(\psi^{\dagger \alpha}_i\psi_{i+1,\alpha}
+\psi^{\dagger \alpha}_{i+1}\psi_{i,\alpha}\right)+H_K.
\label{eq:tbh}
\end{equation}
Here $H_K$ describes the coupling to the impurity spin. For the bulk of
our results we
consider a single $s=1/2$ impurity spin at
the left end of an otherwise open chain. In this case the coupling
to the impurity spin described by $H_K$ takes the form
\begin{eqnarray}
H_K&=&
J{\bf S}_{\rm imp}\cdot\psi_1^{\dagger \alpha}
\frac{\bbox{\sigma}^\beta_\alpha}
{2}\psi_1^\beta\nonumber\\
&=&\frac{1}{2}J\left(\left(\psi^\dagger_{1 \uparrow}\psi_{1 \uparrow}-
\psi^\dagger_{1 \downarrow}\psi_{1 \downarrow}\right)S^z_{\rm imp}+
\psi^\dagger_{1 \uparrow}\psi_{1 \downarrow}S^-_{\rm imp}
+\psi^\dagger_{1 \downarrow}\psi_{1 \uparrow}S^+_{\rm imp}\right).
\label{eq:hk1}
\end{eqnarray}
We also briefly consider the case of two $s=1/2$ impurities located at either
end of the chain, correspondingly $H_K$ becomes:
\begin{equation}
H_K=
J{\bf S}_1\cdot\psi_1^{\dagger \alpha}\frac{\bbox{\sigma}^\beta_\alpha}
{2}\psi_1^\beta+
J{\bf S}_L\cdot\psi_L^{\dagger \alpha}\frac{\bbox{\sigma}^\beta_\alpha}
{2}\psi_L^\beta.
\label{eq:hk2}
\end{equation}
In all our results below we have used $t=1$.

\section{numerical method}\label{sec:num}
We use the density matrix renormalization group (DMRG) method as
developed by S.~R.~White and R.~M.~Noack~\cite{white}. For a detailed
explanation of the method we refer the reader to
Ref.~\onlinecite{white}. The method
is centered around calculating the density matrix and its
corresponding eigenvalues and eigenvectors for two appropriately
defined parts of the total system. Usually this is done by simply
considering the system as having a left and a right part. The eigenvalues
of the density matrices can then be interpreted as the probability for
the subsystem to be in an eigenstate of the density matrix
given the constraint that the total system is in a fixed ``pure"
eigenstate (usually the ground-state). For stable fixed points of
the DMRG method
it can be shown~\cite{stellan} that states in the thermodynamic limit
are well represented by ``matrix product ground
states"~\cite{accardi,fannes,klumper}.

In an iteration
the density matrices for each half of the system contain $4\times m$ states
(including the $S=1/2$ impurity spin). Of these states $m$ are kept
to start the next iteration. We have used $m$ in the range
128-200. It is also extremely useful to use all the symmetries
of the original Hamiltonian and constrain the whole calculation to
a subspace defined by suitable quantum numbers. This increases the
precision of the method dramatically.
{}For the Kondo Hamiltonian that we
consider here we have in addition to the total z-component
of the spin, $S^z_T$ (including both the impurity and electron spin),
and the number of electrons also parity, $P$, for reflection
around the midpoint of the chain and, $RPH$ a particle hole symmetry combined
with a rotation that changes the sign of the z-component of
the spin. The first two are diagonal in the usual
product basis and are thus trivial to implement, the $P$ and $RPH$ symmetries
are non diagonal and a considerable effort has to be expended to implement
these symmetries. The parity $P$, is standard and takes the two values
$1,-1$. This is only applicable when we consider two impurities since
the one impurity model is not symmetric with respect to a reflection
around the middle of the chain.
The $RPH$ symmetry is an on-site symmetry and can be used both for the one
and two impurity model that we consider. It is defined by
\begin{eqnarray}
RPH&:&\ \psi_{j\alpha} \rightarrow\ (-1)^j\psi_j^{\dagger\beta}
\left(\sigma^x\right)_{\beta\alpha}\nonumber\\
RPH&:&\ \psi_{j}^{\dagger \alpha} \rightarrow \
(-1)^j \left(\sigma^x\right)^{\alpha \beta}\psi_{j \beta}\nonumber\\
RPH&:&\ |0>\ \rightarrow\ |F>.
\label{eq:rph}
\end{eqnarray}
Here $|0>$ is the empty state and $|F>$ the completely filled
state. This is an exact symmetry of the Hamiltonian and it
commutes with $P$. We can then specify a state by the
four quantum numbers, the filling factor, $S^z_T$, $P$, and
$RPH$. In the following we shall always work at half-filling,
but we chose the remaining quantum numbers so as to select
appropriate states.
{}For calculations performed on a system with an impurity at both
ends we can use all four quantum numbers, for the case with
only one impurity the chain has no longer reflection symmetry around
the midpoint of the chain and $P$ is no longer a good quantum number.

The bulk of our results are obtained for the case of only one
impurity.
In order to obtain optimal precision
it is necessary to
use a combination of the so called ``infinite" and ``finite" length
DMRG methods~\cite{white}. At each step of the infinite chain method
a complete run of the finite length method is performed.
This is done in the following way. We start by considering a 2 site
system in addition to the impurity spin. Using the infinite lattice method
we generate a 4 site system with matrices representing the impurity
spin plus the 2 left sites of the chain
and another matrix representing the two
right sites. We denote this by $2_L+2_R$, where the impurity spin
is included in the matrix $2_L$. The finite system method
is then used to arrive at a system consisting of $1_L+3_R$ where the
exact $8\times 8$ matrix is used for the impurity spin and first
site for the matrix $1_L$.
At this step the expectation values $<S^z_{\rm imp}>$
and $<S^z_1>$ are calculated since these matrices are known exactly.
The finite lattice method is then used to generate $2_L+2_R$ and
$3_L+1_R$, at each step calculating the expectation value of
the electron spins at the sites where the matrices are known exactly,
i.e. at sites 2,3 and 3,4 respectively.
Then we use the finite system method to generate $3_L+3_R$
and the whole procedure is
repeated. For large chain lengths this is exceedingly slow, but we have
been able to treat chain lengths of up to 50 sites. This method
is exact out to the point where the matrices $N_L$ and $N_R$ have to be
truncated and it has the great merit of yielding the same precision
for the expectation value of the spin components in the middle and
at the end of the chain. This latter point is essential to our
analysis of how the impurity spin is screened at each chain length.

In all cases below we shall take $t=1$ and we always take $L$ even.
{}Furthermore, we shall usually work in the ground-state subspace which
for 1 impurity is given by the quantum numbers, $S^z_T=1/2$, $RPH=1$.
{}For 2 impurities the ground-state subspace is defined by, $S^z_T=0$,
$RPH=1$, and $P=-1$.

\section{Free Chain}\label{sec:free}

Let us first consider the tight binding model in the absence of any
impurity.
\begin{equation}
H_{\rm free}= -t\sum_{i=1}^{L-1}\left(\psi^{\dagger \alpha}_i
\psi_{i+1,\alpha}
+\psi^{\dagger \alpha}_{i+1}\psi_{i,\alpha}\right).
\label{eq:hfree}
\end{equation}
Here the subscript ``free" denotes the free chain. Since we shall
be concerned mainly with free boundary conditions for the chain we
introduce two ``phantom sites", $0$ and $L+1$, where we
require that $\psi(0)=\psi(L+1)=0$. This model can be
solved by transforming to Fourier space and we obtain, in units
of the lattice spacing, $a$,
\begin{equation}
H_{\rm free}= -2t\sum_k \cos(ak)\psi^{\dagger \alpha}_k\psi_{k,\alpha}
\label{eq:hfreek}
\end{equation}
where the allowed values for the wave vector, k, are
\begin{equation}
k=\frac{\pi n}{L+1},\ \ n=1\ldots L,
\end{equation}
and thus $0< k < \pi$. We have now essentially two decoupled
{}Fermi seas for up and down spin electrons respectively. The energy
is then given by the expression
\begin{equation}
E_{\rm free}= -2
t\left(\sum_{k^\uparrow}^{k^{\uparrow}_F}\cos(ak^\uparrow)+
\sum_{k^\downarrow}^{k^\downarrow_F} \cos(ak^\downarrow)\right).
\label{eq:freerg}
\end{equation}
Here $k^{\uparrow}_F, k^\downarrow_F$ denotes the Fermi wave vector
for up and down spin electrons. We can now choose a simple
representation for the field operators.
\begin{equation}
\psi_j=\sqrt{\frac{2}{L+1}}\sum_k \sin(kj)a_k,\ \ \
\psi_j^\dagger=\sqrt{\frac{2}{L+1}}\sum_k \sin(kj)a_k^\dagger.
\label{eq:fieldop}
\end{equation}
We note that with this definition the field operators obey
the commutation relations
$\{\psi_j,\psi^\dagger_j\}=\delta_{j,l}$.

This solution leads to the interesting fact that for a chain with
an {\it odd} number of sites at exactly half filling, i.e. one electron
per site, the magnetization per site (the expectation value
of the z-component of the electron spin, $<S^z_j>$,
is given by the following expression
\begin{eqnarray}
<S^z_j>&=&
\frac{1}{2}<\psi^{\uparrow \dagger}_j\psi^{\uparrow}_j-
\psi^{\downarrow \dagger}_j\psi^{\downarrow}_j>\nonumber\\
&=&\frac{1}{L+1}\sin^2\frac{\pi j}{2}.
\label{eq:oddsus}
\end{eqnarray}
Thus the on-site magnetization, $<S^z_j>$, is non-zero only on {\it odd}
sites, an artifact due to the open boundary conditions.
This result can be reproduced by the DMRG method.

{}For an even length chain, at half-filling, the on-site
magnetization is always zero in the ground-state where $S_T=0$,
since the expectation value of the electron spin is zero at
every site due to rotation symmetry. However an excited state
with $S_T=1$ will have unpaired spin up electrons in the states
with $k=L\pi/(2L+2), (L+2)\pi/(2L+2)$. Redoing the above calculation
above for an {\it even} length chain in the $S_T=1$ state we obtain:
\begin{equation}
<S^z_j>=\frac{1}{L+1}\left[\sin^2\frac{L\pi j}{2(L+1)}+
\sin^2\frac{(L+2)\pi j}{2(L+1)}\right].
\label{eq:strong}
\end{equation}
Equivalent results can be obtained for higher excited states.

\section{Weak coupling perturbation}\label{sec:weak}
\subsection{$<S^z_j>$}
By considering the term $H_K$ in Eq.~(\ref{eq:tbh}) as a perturbation
we can do first order perturbation theory in $(J/t)$. We start with
a system at half filling with an {\it even} number of sites $L$ and
we shall take the total
z-component of the spin to be $1/2$, $S^z_T=1/2$.
The impurity spin is at the far end of the chain at
site 1. Here we consider only one impurity described by
$H_K$ in Eq.~(\ref{eq:hk1}).
Since we are only considering first order perturbation theory
only the term involving $S^z$ in $H_K$ will  contribute.
Thus, we have
for the unperturbed ground-state
\begin{equation}
\phi_0 = |\phi>_{\rm free}\times |\uparrow>.
\label{eq:freegnd}
\end{equation}
Here $|\phi>_{\rm free}$ is the half-filled sea described in the preceding
section and $|\uparrow>$ denotes the spin up state of the impurity
spin. The first order perturbation to the wave function is
\begin{equation}
|\phi_1>=\frac{1-P}{E_0-H_{\rm free}}\frac{J}{4}
\left(\psi_{1}^{\dagger \uparrow}
\psi_{1}^\uparrow-\psi_{1}^{\dagger \downarrow}\psi_{1}^\downarrow\right)
|\phi>_{\rm free}\times |\uparrow>.
\label{eq:kurt}
\end{equation}
Here $P$ is the projection operator onto the ground-state
Eq.~(\ref{eq:freegnd}).
When $L$ is even all single particle states with energy
$\varepsilon_k= -2t\cos k$ below the Fermi energy at $\varepsilon_F=0,
k_F=\pi/2$ are filled for the unperturbed chain. With this notation
we can rewrite Eq.~(\ref{eq:kurt}) as
\begin{equation}
|\phi_1> = \frac{J}{2L+2}\sum_{k,k', \varepsilon_k<0,\varepsilon_{k'}>0}
\frac{\sin k \sin k'}{\varepsilon_k-\varepsilon_{k'}}
\left(a^\dagger_{k^{\prime} \uparrow}a_{k \uparrow}-
a^\dagger_{k^{\prime} \downarrow}a_{k \downarrow}\right)
|\phi>_{\rm free}\times
|\uparrow>.
\end{equation}
Thus, the on-site magnetization becomes
\begin{equation}
<S^z_j> =
\frac{2J}{(L+1)^2}\sum_{k,k^{\prime},
\varepsilon_k<0,\varepsilon_{k^{\prime}}>0}
\frac{\sin kj\sin k^{\prime}j \sin k
\sin k^{\prime}}{\varepsilon_k-\varepsilon_{k^{\prime}}}.
\label{eq:sz1D}
\end{equation}
If we now set $k^{\prime}=\pi-k^{\prime\prime}$ we see that
$\varepsilon_{k^{\prime}}=-\varepsilon_{k^{\prime\prime}},
\sin k^{\prime}j = -(-1)^j\sin k^{\prime\prime}j$
and we can therefore rewrite the above equation as
\begin{equation}
<S^z_j>=\frac{J}{t}\frac{(-1)^j}{(L+1)^2}
\sum_{n,m=1}^{L/2}\frac{\sin k_n\sin k_m \sin k_nj \sin k_mj}
{\cos k_n+\cos k_m},
\label{eq:1sus}
\end{equation}
where as above we have $k_n=\pi n /(L+1)$.
{}For $j\gg 1$ Eq.~(\ref{eq:1sus}) can be analytically evaluated
as:
\begin{equation}
<S^z_j> \to {(J/t) \over 4\pi j}(-1)^j.
\label{eq:1susasymp}
\end{equation}
This formula is basically the 1D version of the RKKY formula,
with $1/r^3$ replaced by $1/r$ for trivial dimensional reasons.
Note that this expression only has a staggered part, not a uniform
part.  The uniform part is $O(J^2)$.  It is quite easy to see
that the sum over all j of Eq.~(\ref{eq:1sus}) vanishes exactly.
(This is simply a consequence of the fact that the total electron
spin of the unperturbed ground-state, with L even, is zero.)

\begin{figure}
\centering
\epsfysize=12 cm
\leavevmode
\epsffile{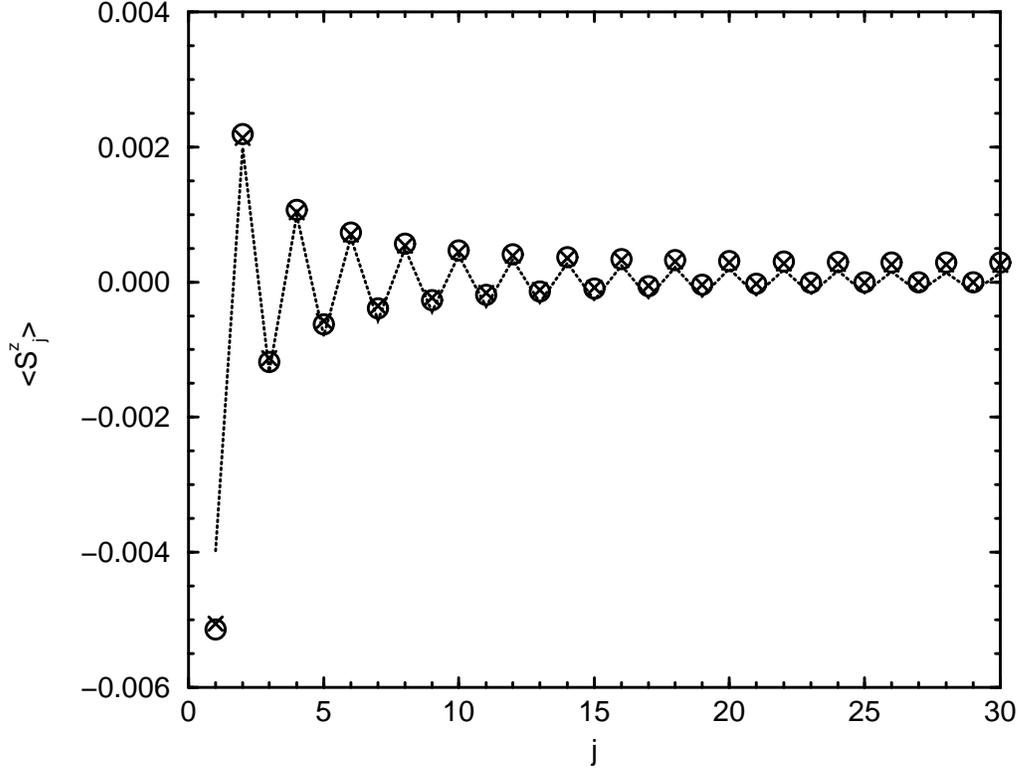}
\caption{The z-component of the electron spin as a function of
site index j, for a chain of length $L=30$. At the left end of the
chain is an $S=1/2$ impurity spin. The Kondo coupling is $J=0.05$.
The crosses indicate
the first order perturbation result, Eq.~(\protect\ref{eq:1sus}), the
circles denote DMRG results for the state $S^z_T=1/2, RPH = 1.$
$m=128$ states were kept.
The dotted line is the asymptotic result Eq.~(\protect\ref{eq:1susasymp}).
}
\label{fig:szj.j0.05}
\end{figure}

The above result, Eq.~(\ref{eq:1sus}), can be compared to DMRG
results obtained for weak couplings. In Fig.~\ref{fig:szj.j0.05} we
show results for a 30 site system with one impurity at the left end.
The circles denote $<S^z_j>$ calculated with the DMRG method, while
the crosses are Eq.~(\ref{eq:1sus}). The dotted line indicates
the asymptotic form Eq.~(\ref{eq:1susasymp}).
The calculation was performed keeping
$m=128$ states in the ground-state subspace defined
by $RPH=1$, $S^z_T=1/2$. Clearly there is a very good agreement
between the perturbative results, Eq.~(\ref{eq:1sus}), and the DMRG results.

\begin{figure}
\centering
\epsfysize=12 cm
\leavevmode
\epsffile{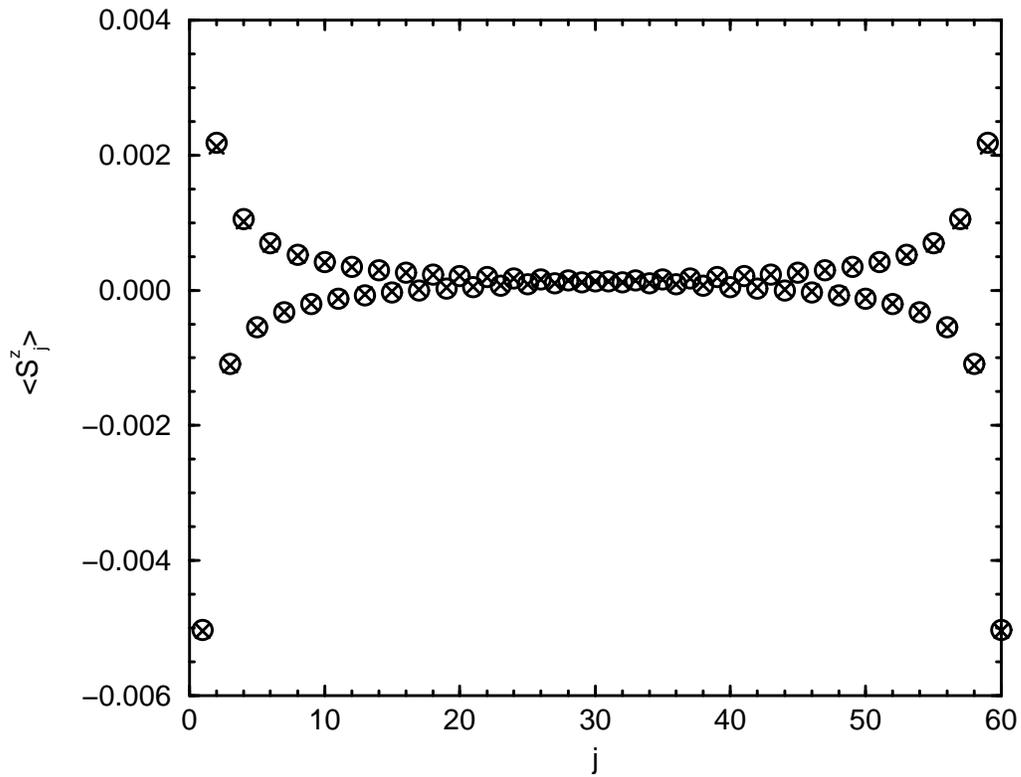}
\caption{The z-component of the electron spin as a function of
site index j, for a chain of length $L=60$. Impurity spins are
present at both ends of the
chain. The Kondo coupling is $J=0.05$.
The crosses indicate
the first order perturbation result, Eq.~(\protect\ref{eq:2chi}), the
circles denote DMRG results for the state $S^z_T=1, RPH = 1, P=1.$
}
\label{fig:weak}
\end{figure}

Equation~(\ref{eq:1sus}) can also be compared to results obtained using
the DMRG method with two impurities, one at either end of the chain.
In Fig.~\ref{fig:weak} we show the expectation value
of the z-component of the electron spin as a function of chain index, j,
for a Kondo coupling, $J$, of 0.05.
In this calculation we have a $S=1/2$ impurity spin at each end of the
chain. The total z-component of the electron and impurity spin was
chosen to be $S^z_T=1$ so as to polarize the two impurity spins as much
as possible, $m=150$ states were kept and we fixed the two remaining quantum
numbers $RPH=1, P=1$. As usual we work at half filling. The DMRG results
are shown as the circles in Fig.~\ref{fig:weak}. In order to compare
these results to Eq.~(\ref{eq:1sus}) we must sum the contribution from
the impurity at both ends of the chain. If we denote by $S_1^{z (1)}$
the result of Eq.~(\ref{eq:1sus}) for the contribution for one impurity
spin we obtain
\begin{equation}
S_j^{z (2)}=S_j^{z (1)}+S_{L-j+1}^{z (1)}.
\label{eq:2chi}
\end{equation}
This expression is shown as the $\times$ in Fig.~\ref{fig:weak}.
As clearly seen in Fig.~\ref{fig:weak} there is an excellent agreement
between the perturbation results and the DMRG results for the chain length,
$L=60$, considered.

\subsection{$S^z_{\rm imp}$}
The expectation value of $S^z_{\rm imp}$ can also be
evaluated in first order perturbation theory.
Using Eq.~(\ref{eq:kurt}) and remembering to include the contributions
from the x and y parts of Kondo interaction, we obtain:
\begin{equation}
<S^z_{\rm imp}> \approx 1/2 - \left[{J\over
(L+1)}\right]^2\sum_{\epsilon_k<0,
\epsilon_{k'}>0}\left[{\sin k\sin k'\over
\epsilon_k-\epsilon_{k'}}\right]^2.
\label{eq:simpweak}
\end{equation}
Where we must be careful to include the correction to
the ground-state wave-function normalization.
In the large L limit, replacing the
sum by an integral, the integral has a log divergence.  Thus
the last term goes like $\ln L$.

\begin{figure}
\centering
\epsfysize=12 cm
\leavevmode
\epsffile{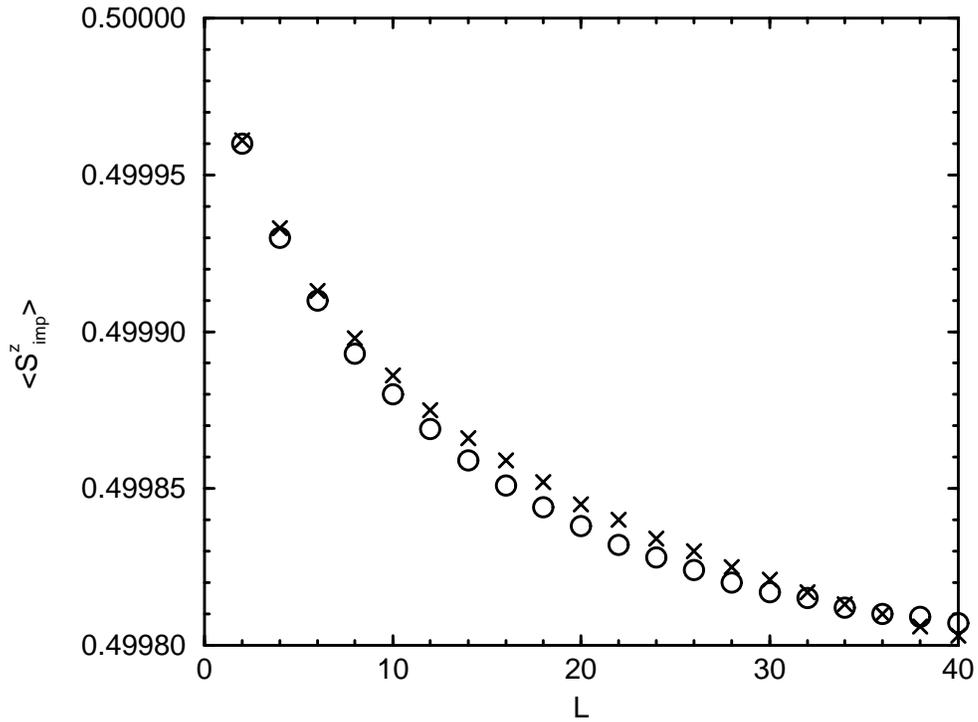}
\caption{The z-component of the impurity spin as a function
of chain length, $L$.
One impurity is present at the left end of the chain.
The Kondo coupling is $J=0.05$.
The crosses indicate
the first order perturbation result, Eq.~(\protect\ref{eq:simpweak}), the
circles denote DMRG results for the state $S^z_T=1/2, RPH = 1.$
}
\label{fig:szimp}
\end{figure}

In Fig.~\ref{fig:szimp} we show DMRG results for
the z-component of the impurity spin as a function
of chain length, $L$.
The results shown are for a very weak Kondo coupling, $J=0.05$.
One impurity is present at the left end of the chain.
The crosses indicate
the first order perturbation result, Eq.~(\ref{eq:simpweak}), the
circles denote DMRG results for the state $S^z_T=1/2, RPH = 1.$
A good agreement between the perturbative and numerical results
is evident.

\subsection{Correlation Function}

\begin{figure}
\centering
\epsfysize=12 cm
\leavevmode
\epsffile{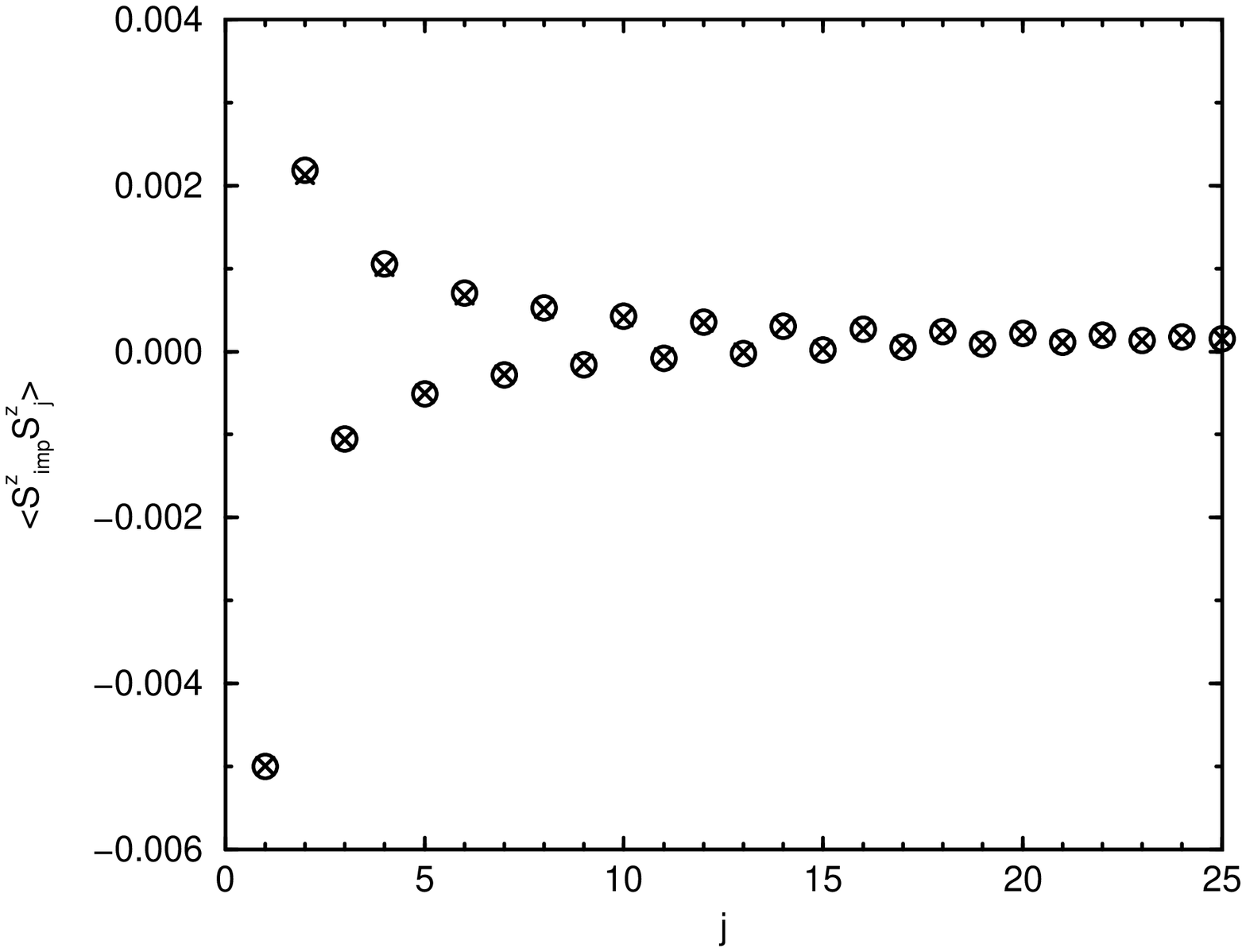}
\caption{The correlation function $<S^z_{\rm imp}S^z_j>$ as a function of
site index j (circles), for a chain of length $L=50$. At both ends of the
chain are $S=1/2$ impurity spins. Only {\it half} the chain is
shown. The Kondo coupling is $J=0.050$,
$m=150$ states are kept with $RPH=1, P=1$,  and $S^z_T=1$. The circles
denote the DMRG results.
The crosses indicate the perturbation result, 1/2
times Eq.~(\protect\ref{eq:1sus}).
}
\label{fig:simpszi.j0.05}
\end{figure}

It is also straightforward to obtain the correlation function
$<S^z_{\rm imp}S^z_j>$ to first order in $J/t$. It is simply
$1/2$ times Eq.~(\ref{eq:2chi}). In Fig.~\ref{fig:simpszi.j0.05}
we show DMRG results for $<S^z_{\rm imp}S^z_j>$ (circles) for a
50 site chain with impurities at both ends. The Kondo coupling
is very weak $J=0.05$. Note that only
{\it half} the chain is shown. The calculation has been performed
with $m=150$ states in the subspace $RPH=1, P=1$, and $S^z_T=1$.
The crosses denote the results from Eq.~(\ref{eq:1sus})
multiplied by 1/2. We see an excellent agreement between the
DMRG and the perturbative results.

\section{Strong Coupling}\label{sec:strong}

\subsection{Energy}
In the $J=\infty$ limit the impurity traps an electron and
forms a tightly bound singlet leaving a free chain
with $L-1$ sites. For finite $J$ the singlet can be polarized
and we can do perturbation theory in the hopping term between
the first and second site. We thus take
\begin{eqnarray}
H&=&
J{\bf S}_{\rm imp}\cdot\psi_i^{\dagger \alpha}
\frac{\bbox{\sigma}^\beta_\alpha}
{2}\psi_i^\beta+H_{\rm free}(L-1) + V,\\
V&=&
-t\left(\psi^{\dagger \alpha}_1\psi_{2,\alpha}
+\psi^{\dagger \alpha}_{2}\psi_{1,\alpha}\right).
\label{eq:hstrong}
\end{eqnarray}
Here $H_{\rm free}$ is the free chain Hamiltonian Eq.~(\ref{eq:hfree})
for $L-1$ sites.
In the following we shall regard the hopping term, $V$, between site 1 and 2
as the perturbation.

\begin{figure}
\centering
\epsfysize=7 cm
\leavevmode
\epsffile{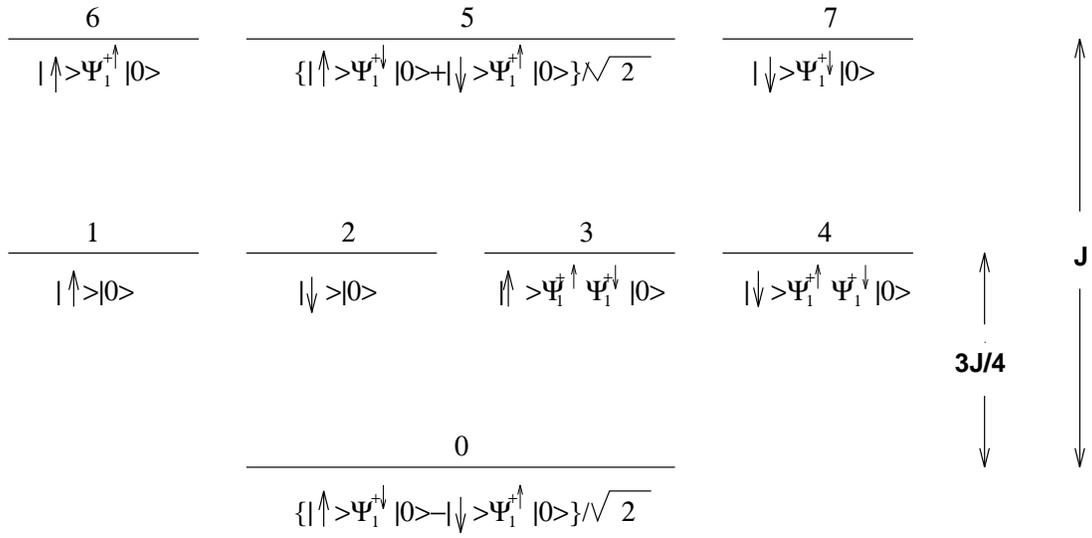}
\caption{The 8 states for the impurity site.
}
\label{fig:levels}
\end{figure}

The strongly bound singlet on the first site of the chain can be
excited into any of the seven excited states shown in
{}Fig.~\ref{fig:levels}. The excited states form a quadruplet with energy
$3J/4$ and a triplet with energy $J$ relative to the ground-state.
The perturbation, $V$, has only non-zero matrix elements between the
ground-state and the quadruplet. Calculating the partial matrix elements
in the impurity part of the Hilbert space we find
\begin{eqnarray}
V_{1 0}&=<1|V|0> =& \frac{1}{\sqrt{2}}\psi_2^{\dagger \downarrow}\nonumber\\
V_{2 0}&=<2|V|0> =& -\frac{1}{\sqrt{2}}\psi_2^{\dagger \uparrow}\nonumber\\
V_{3 0}&=<3|V|0> =& -\frac{1}{\sqrt{2}}\psi_2^{\uparrow}\nonumber\\
V_{4 0}&=<4|V|0> =& -\frac{1}{\sqrt{2}}\psi_2^{\downarrow}.
\label{eq:melements}
\end{eqnarray}
{}Following standard second order perturbation theory we can now
calculate the energy shift due to the perturbation $V$. If we denote by
$|0>|F>$ the unperturbed ground-state composed of the singlet
and the free chain we find
\begin{eqnarray}
\Delta E&=& -\frac{4t^2}{3J}\sum_{\alpha=1}^4<F|<0|V|\alpha><\alpha|V|0>|F>
\nonumber\\
&=&-\frac{4t^2}{6J}<F|\{\psi_2^{+ \alpha},\psi_{2 \alpha}\}|F>
\nonumber\\
&=&-\frac{4t^2}{3J}.
\label{eq:De}
\end{eqnarray}
We thus find for the total energy for 1 and 2 impurities, respectively,
\begin{eqnarray}
E_{\rm 1\ impurity}&=&-\frac{3}{4}J+E_{\rm free}(L-1)-\frac{4t^2}{3J}
\nonumber\\
E_{\rm 2\ impurity}&=&-\frac{6}{4}J+E_{\rm free}(L-2)-\frac{8t^2}{3J}.
\label{eq:estrong}
\end{eqnarray}
These perturbation results compare favorably with the DMRG results
at sufficiently strong coupling.

\subsection{wave-function}
We can now calculate the wave-function to second order in perturbation
theory. We find
\begin{equation}
|\phi>=|0>|F>-\frac{4t}{3J}\sum_{\alpha=1}^4V_{\alpha 0}|\alpha>|F>
+\frac{4}{3}(\frac{t}{J})^2\sum_{\alpha=1}^4V_{5 \alpha}V_{\alpha 0}|5>|F>
+\cdots.
\label{eq:wfctn}
\end{equation}
{}For the calculations we shall consider here these are the only terms
that will contribute. We need to calculate $V_{\alpha 5}$.
We find for the partial matrix elements in the impurity part of the Hilbert
space,
\begin{eqnarray}
V_{1 5}&=<1|V|5> =& \frac{1}{\sqrt{2}}\psi_2^{\dagger \downarrow}\nonumber\\
V_{2 5}&=<2|V|5> =& \frac{1}{\sqrt{2}}\psi_2^{\dagger \uparrow}\nonumber\\
V_{3 5}&=<3|V|5> =& -\frac{1}{\sqrt{2}}\psi_2^{\uparrow}\nonumber\\
V_{4 5}&=<4|V|5> =& \frac{1}{\sqrt{2}}\psi_2^{\downarrow}.
\label{eq:m5elements}
\end{eqnarray}
We can now proceed to evaluate the expectation
value of $S^z_1$, $S^z_2$, $S_{\rm imp}^z$ in perturbation theory.

\subsection{$S^z_j$, $S_{\rm imp}^z$}\label{sec:sss}

We begin by considering the z-component of the electron spin
on the first site, $S^z_1$. In the unperturbed system $S^z_1$ must
be zero since it is locked in a singlet with the impurity spin.
The first non-zero contribution to $S^z_1$ is second order in $t/J$.
Since
\begin{equation}
S^z_1|0> = -\frac{1}{2}|5>,
\end{equation}
we find that only the second term in Eq.~(\ref{eq:wfctn}) will
contribute and we get
\begin{equation}
<S^{z(2)}_1>=2<F|<0|S^z_1\frac{4}{3}(\frac{t}{J})^2\sum_{\alpha=1}^4V_{5
\alpha}V_{\alpha 0}|5>|F>=-\frac{4}{3}(\frac{t}{J})^2
<F|\psi_2^{\dagger \uparrow}\psi_2^{\uparrow}-\psi_2^{\dagger
\downarrow}\psi_2^{\downarrow}|F>^{(2)}.
\label{eq:s1z}
\end{equation}
{}From this we can derive two results; if the state
{}$|F>$ describes an unperturbed chain with $L-1$ sites we find
from Eq.~(\ref{eq:oddsus})
\begin{equation}
<S^{z(2)}_1>=-\frac{8}{3}(\frac{t}{J})^2\frac{1}{L},
\end{equation}
since in that case the matrix element in Eq.~(\ref{eq:s1z}) is
simply $2/L$. We can also consider the case where $|F>$ describes
a free chain with $L-2$ sites but now in a state with
$S^z_T=1$. This is convenient for comparing with results
with 2 impurities where 2 sites will be quenched out.
It is in that case convenient to work in the state with $S^z_T=1$.
We then find
\begin{equation}
<S^{z(2)}_1>=-\frac{8}{3}(\frac{t}{J})^2\frac{1}{L-1}
[\sin^2\frac{\pi
(L-2)}{2(L-1)}+\sin^2\frac{\pi L}{2(L-1)}].
\label{eq:sz1}
\end{equation}

\begin{figure}
\centering
\epsfysize=12 cm
\leavevmode
\epsffile{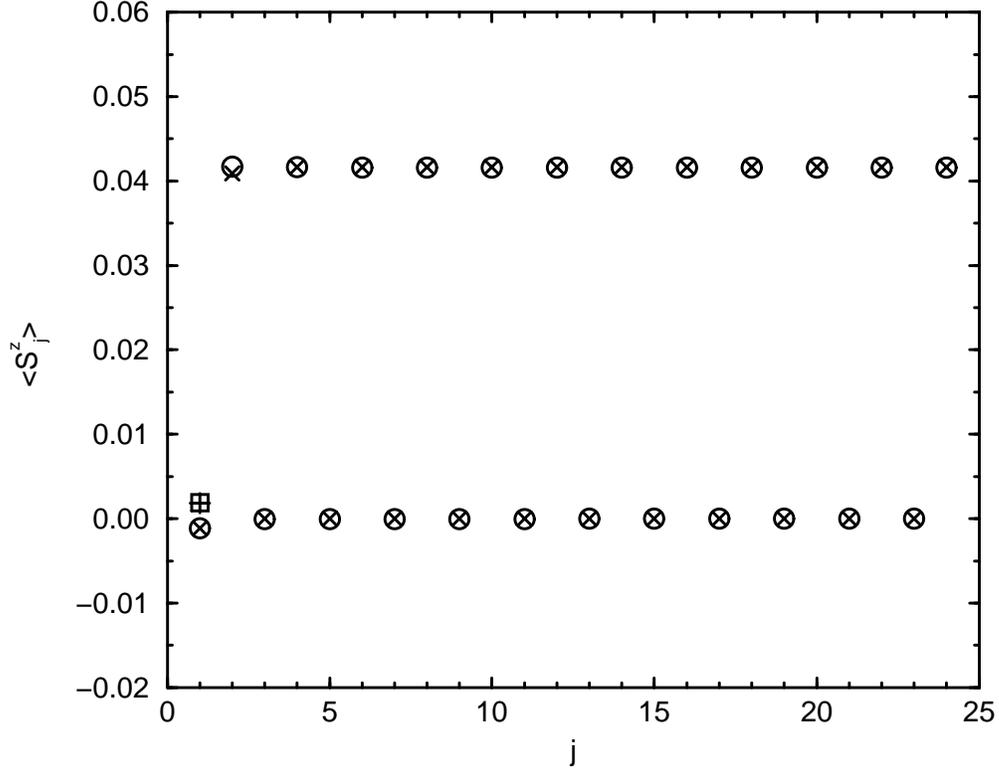}
\caption{The z-component of the electron spin as a function of
site index j (circles), for a chain of length $L=24$. At the left end of the
chain is a $S=1/2$ impurity spin. The Kondo coupling is $J=10$,
$m=128$ states are kept with $RPH=1$, $S^z_T=1/2$. The square
denote the DMRG result for the impurity spin.
The crosses indicate
the perturbation result,
Eqs.~(\protect\ref{eq:oddsus}), (\protect\ref{eq:sz1})
and (\protect\ref{eq:sz2}). The plus is the result for $<S_{\rm imp}>$
from Eq.~(\protect\ref{eq:simp}).
}
\label{fig:j10}
\end{figure}

In a similar fashion $S_{\rm imp}^{z(2)}$ can be calculated to
second order in $t/J$. We first consider the case with one impurity.
Again $S^z_{\rm imp}$ is zero in
the unperturbed state. For $S^z_{\rm imp}$ we find
\begin{equation}
S^z_{\rm imp}|0>=\frac{1}{2}|5>.
\end{equation}
Thus we get the same term as before but with a different sign. However,
since $S^z_{\rm imp}$ is non-zero in the quadruplet we get an
additional term from the second term in Eq.~(\ref{eq:wfctn}).
We then have
\begin{eqnarray}
<S^{z(2)}_{\rm imp}>&=&\frac{16}{9}(\frac{t}{J})^2
\sum_{\alpha=1}^4<F|V_{0\alpha}V_{\alpha 0}|F>
<\alpha|S^z_{\rm imp}|\alpha>
+\frac{4}{3}(\frac{t}{J})^2
<F|\psi_2^{\dagger \uparrow}\psi_2^{\uparrow}-\psi_2^{\dagger
\downarrow}\psi_2^{\downarrow}|F>\nonumber\\
&=&\frac{20}{9}(\frac{t}{J})^2
<F|\psi_2^{\dagger \uparrow}\psi_2^{\uparrow}-\psi_2^{\dagger
\downarrow}\psi_2^{\downarrow}|F>^{(2)}.
\label{eq:simpz}
\end{eqnarray}
As before we can now obtain the results for
the case where $|F>$ describes a chain with $L-1$ sites,
ie when only one impurity is present.
\begin{equation}
<S^{z(2)}_{\rm imp}>=\frac{40}{9}(\frac{t}{J})^2\frac{1}{L}.
\label{eq:simp}
\end{equation}
And equivalently for the state with $L-2$ sites and $S^z_T=1$
where we have the 2 impurity case in mind.
\begin{equation}
<S^{z(2)}_{\rm imp}>=\frac{40}{9}(\frac{t}{J})^2\frac{1}{L-1}
[\sin^2\frac{\pi
(L-2)}{2(L-1)}+\sin^2\frac{\pi L}{2(L-1)}].
\end{equation}
To this order in perturbation theory we have in addition the
following equality for the second order contribution to
$<S^z_2>$:
\begin{equation}
<S_2^{z(2)}>= -<S_1^{z(2)}>-<S^{z(2)}_{\rm imp}>.
\label{eq:sz2}
\end{equation}
Note that for $<S^z_2>$ we still have the term $1/L$
from Eq.~(\ref{eq:oddsus})
which we can apply here at strong coupling.
These equations are obeyed to a high accuracy at sufficiently high
coupling $J$. To illustrate this we show in Fig.~\ref{fig:j10} results
for one impurity with a Kondo coupling of $J=10$ for $L=24$.
The calculation is done
keeping $m=128$ states with $RPH=1$ and $S_T^z=1/2$. The circles
denote the DMRG results
for $<S^z_j>$ and the square denote $<S^z_{\rm imp}>$.
The crosses denote the results from Eq.~(\ref{eq:oddsus}) for a
23 site chain with the corrections from Eqs.~(\ref{eq:sz1})
and (\ref{eq:sz2}). The plus is the result for $<S^z_{\rm imp}>$
from Eq.~(\ref{eq:simp}). From the results in Fig.~\ref{fig:j10}
we find excellent agreement between the above perturbation results
and our DMRG results at strong coupling.

We note that the fact that $<S^z_j> \to {\rm constant}/L$ for
$j\gg \xi_K$, and thus in a sense is longer range than RKKY (1/r),
is analogous to
the prediction that $\chi(r,T)$ should be longer range when
the screening cloud has formed (see discussion below Eq.~(\ref{chilt})),
than when it has not.

\subsection{correlation function}
It is also rather straight forward to calculate the correlation
function $<S^z_{\rm imp}S^z_j>$ to second order in $t/J$. Again
the unperturbed result is just zero for $j>1$. Let us first consider the
case where $j>2$, in which case we get from
Eq.~(\ref{eq:simpz})
\begin{eqnarray}
<S^z_{\rm imp}S^z_j>^{(2)}&=&
\frac{20}{9}(\frac{t}{J})^2<F|S^z_j(\psi_2^{\dagger
\uparrow}\psi_2^{\uparrow}-\psi_2^{\dagger
\downarrow}\psi_2^{\downarrow})|F>\nonumber\\
&=&\frac{10}{9}(\frac{t}{J})^2 G(L-2,j-1) \ \ j>2.
\end{eqnarray}
Here we consider the case where $|F>$ describes a free chain with
$L-2$ sites corresponding to two impurities, and $G$ is given by
\begin{equation}
G(L,j)=<F|
(\psi_1^{\dagger \uparrow}\psi_1^{\uparrow}-
\psi_1^{\dagger \downarrow}\psi_1^{\downarrow})
(\psi_j^{\dagger \uparrow}\psi_j^{\uparrow}-
\psi_j^{\dagger \downarrow}\psi_j^{\downarrow})
|F>.
\end{equation}
{}For the states describing the free chain $G$ can easily be calculated.
{}For the state that we consider here with $L-2$ sites and $S^z_T=0$,
we find
\begin{eqnarray}
G(L,j)&=&\frac{8}{(L+1)^2}(\sum_{k<k_F}\sin k\sin kj)
(\sum_{k>k_F}\sin k\sin kj)\nonumber\\
 &=& -\frac{8}{(L+1)^2}(\sum_{k<k_F}\sin k\sin kj)^2.
 \label{eq:glj}
\end{eqnarray}
As before we have $k=\pi n/(L+1),\ n=1\cdots L$.
When $j=2$ we have to be somewhat more careful. Reanalyzing the
two terms in Eq.~(\ref{eq:simpz}) we find that due to the
non-commutativity
of $S^z_2$ and $V$ we effectively get a sign change on the first
term and thus
\begin{equation}
<S^z_{\rm imp}S^z_2>^{(2)}=\frac{2}{9}(\frac{t}{J})^2G(L-2,1)=
\frac{1}{9}(\frac{t}{J})^2.
\label{eq:glj2}
\end{equation}
{}Furthermore we must have that
\begin{equation}
<S^z_{\rm imp}S^z_1>^{(2)}=-\frac{1}{4}-\frac{\Delta E}{3J},
\label{eq:ss1}
\end{equation}
for the case where we have only one impurity where $\Delta E$
is given by Eq.~(\ref{eq:De}).
Except for the first argument to $G$ we note that the above
results for the correlation function
do not depend in any essential way on whether we consider
1 or 2 impurities.

\begin{figure}
\centering
\epsfysize=12 cm
\leavevmode
\epsffile{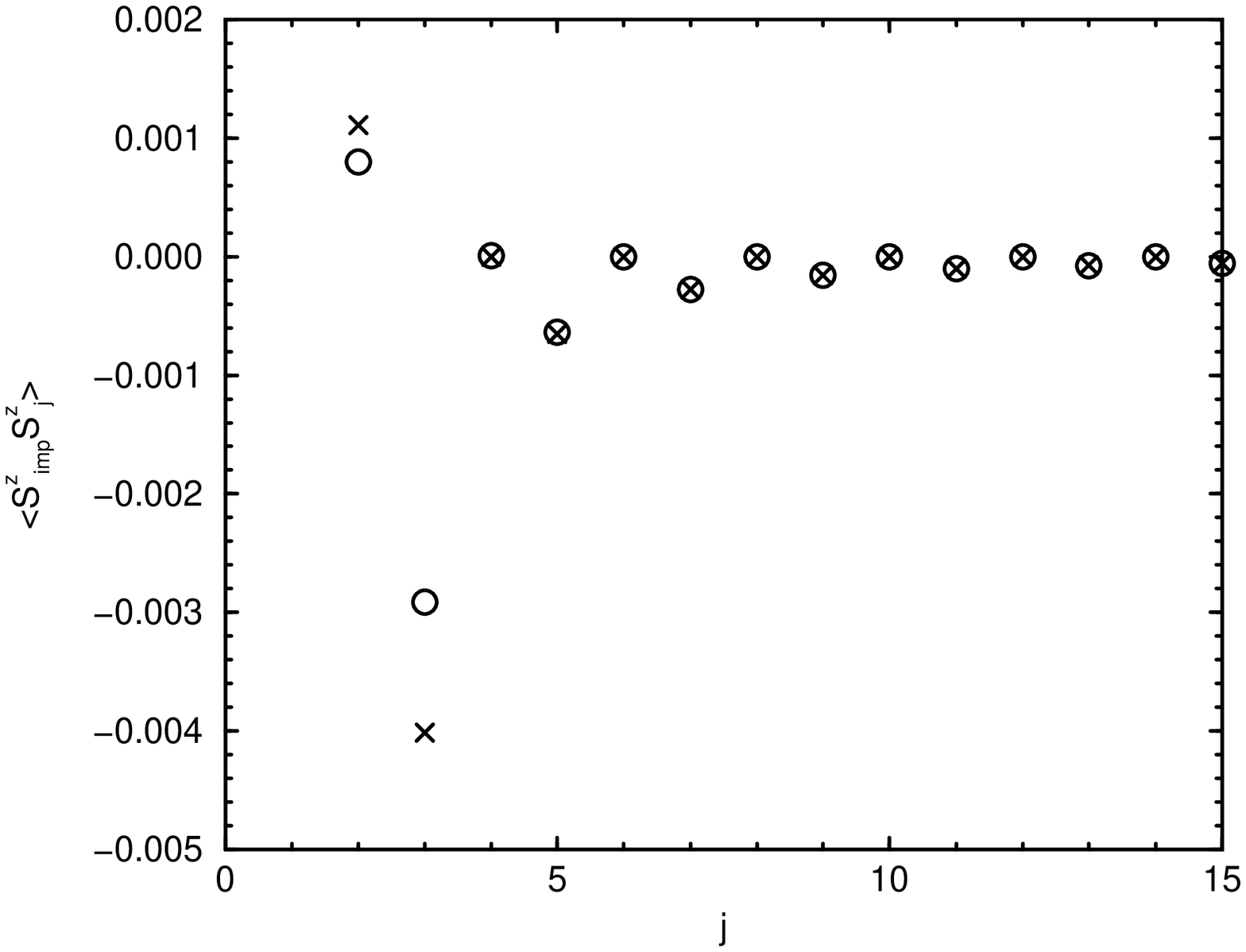}
\caption{The correlation function $<S^z_{\rm imp}S^z_j>$ as a function of
site index j (circles), for a chain of length $L=30$. At both ends of the
chain are $S=1/2$ impurity spins. Only {\it half} the chain is
shown. The Kondo coupling is $J=10$,
$m=150$ states are kept with $RPH=1, P=-1$,  and $S^z_T=0$. The circles
denote the DMRG results.
The crosses indicate
the perturbation result,
Eqs.~(\protect\ref{eq:glj}) and (\protect\ref{eq:glj2}).
}
\label{fig:simpszi.j10}
\end{figure}

We illustrate this with a calculation of $<S_{\rm imp}^zS^z_j>$
for a 30 site chain with an impurity at both ends. We keep
$m=150$ states and work in the ground-state subspace with
$RHP=1, P=-1$, and $S^z_T=0$. We first consider
$<S^z_{\rm imp}S^z_1>$. From the DMRG we find
$<S^z_{\rm imp}S^z_1>=-0.24636$. This can be compared with
the result from Eq.~(\ref{eq:ss1}) which gives
$<S^z_{\rm imp}S^z_1>^{(2)}=-.24555$ in good agreement. The
results for $j\geq 2$ are shown in Fig.~\ref{fig:simpszi.j10}.
The circles denote the DMRG results and the crosses the results from
Eqs.~(\ref{eq:glj}) and (\ref{eq:glj2}). Clearly the discrepancy
is largest for $j=2,3$. As a function of chain length, $L$, we
have observed that this discrepancy decreases. Over all the agreement
with the perturbative results is very good.

\section{Crossover Regime}\label{sec:cross}
Having checked that the DMRG method yields the expected perturbation
results in the strong and weak coupling limit we now proceed
to study the crossover behavior and the scaling predicted by
Eqs.~(\ref{scaling}).
Since the numerical results are all obtained
at $T=0$ for finite systems, where $S^z_{\rm tot}$ is a constant,
the role played by $\Delta \chi \equiv \chi(r,T,J)-\rho/2$ is taken by the
the {\it on-site magnetization}, $\Delta S \equiv <S^z_j>$
(expectation value of the z-component of the electron spin at site $j$).
Hence, a generalized finite size form of Eqs.~(\ref{scaling})
 should
apply to $<S^z_j>$ at $T=0$. Such a finite-size form, applicable
to the numerical results for finite $L$, is easily obtained;
we  simply substitute $L$ for the
thermal length, $v_F/T$. In this way we get
$rT/v_F\to r/L$ and $T/T_K\to \xi_K/L$.
We note
that in order to apply Eq.~(\ref{scaling}) which was
derived for a susceptibility (Eq.~(\ref{eq:locsus})),
to the on-site magnetization
we need to multiply by $T$. We then obtain:
\begin{equation}
<S^z_j> = {1\over L}\left\{
\widetilde f\left({j\over \xi_K}, {L\over \xi_K}\right)(-1)^j+
\widetilde g \left({j\over \xi_K},
{L\over \xi_K}\right)\right\},
\label{eq:magscaling}
\end{equation}
where we have kept the finite-size equivalent of $\chi_{\rm un}$
as well as $\chi_{2k_F}$.   We note that this form assures that $\sum_j
<S^z_j>\approx  (\xi_K/L)\int dx \tilde g(x,L/\xi_K)$ is a function of
$\xi_K/L$ only.

\subsection{Scaling of $L<S^z_{j}>$}

We now proceed to test the above scaling form for the finite
systems we have been able to study numerically.
We start by considering how the weak and strong coupling results
of Sections~\ref{sec:weak} and \ref{sec:strong} can be cast
into a from consistent with Eq.~(\ref{eq:magscaling}). First we note
that the strong coupling expression Eq.~(\ref{eq:oddsus}) obviously
obeys the scaling form. However, the asymptotic weak coupling result,
Eq.~(\ref{eq:1susasymp}) doesn't seem to obey the scaling form due
to the explicit dependence on $J/t$. Fortunately it is possible to
remedy this by noting that $J/t\sim 1/\ln(\xi_K/L)$.
This we see in the following way:
We can write our 2 scaling
variables as $j/L$ and $\xi_K /L$.  Alternatively, and perhaps better,
we can replace $\xi_K /L$ by the effective renormalized dimensionless
Kondo coupling at scale $L$.  Here we define a dimensionless Kondo
coupling as $\lambda \equiv J/t$.
Combining Eqs. (\ref{lambda_eff}) and (\ref{xi_K})
we get using $\Lambda^{-1}=L$:
\begin{equation}
\lambda_{eff} = {1\over \ln (\xi_K /L)}.
\end{equation}
As $L$ becomes small (ie.
approaches 1) $\lambda_{eff}(L)$ approaches the bare coupling
constant, $\lambda$.  Thus, for small $L$, and weak coupling, we have,
by substituting ${1/ \ln (\xi_K /L)}$ for $(J/t)$:
\begin{equation}
L<S^z_j> \to (-1)^j{L\over j}{1\over 4\pi \ln (\xi_K /L)},
\end{equation}
perfectly consistent with scaling.

\begin{figure}
\centering
\epsfysize=12 cm
\leavevmode
\epsffile{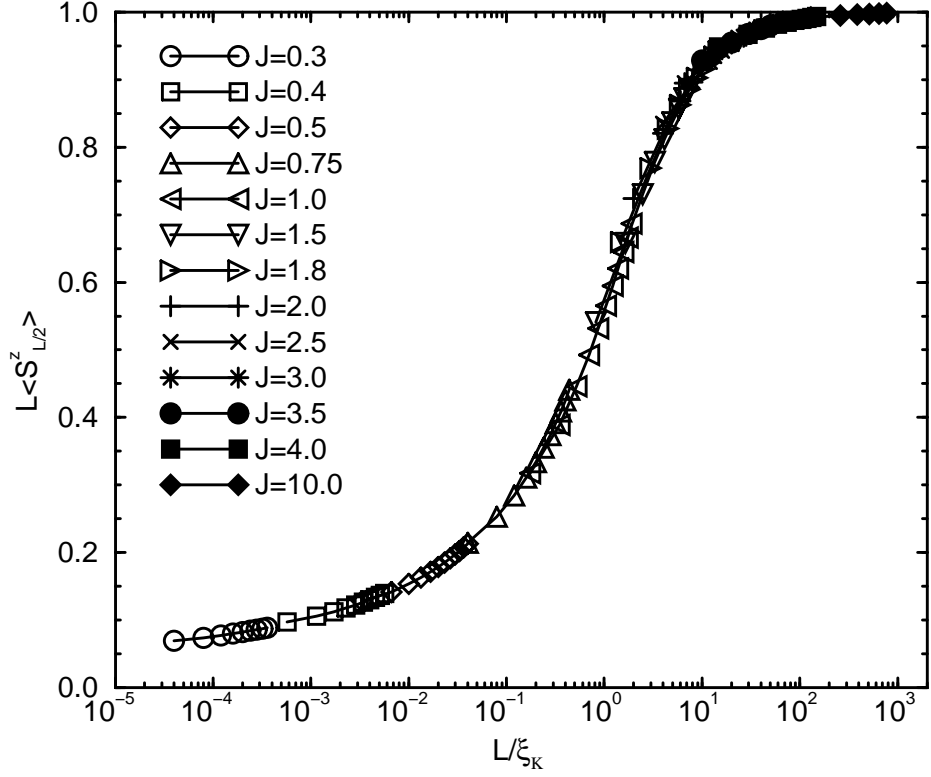}
\caption{Logaritmic plot of $L<S^z_{L/2}>$
as a function of chain length $L/\xi_K$ for
a range of different coupling constants.
The initial point corresponds in
all cases to $L=4$.
The solid lines are guides to the eye.
The strong coupling limit corresponds to $L<S^z_{L/2}>\approx 1$.
}
\label{fig:l2scale}
\end{figure}

{}First we consider $L<S^z_{L/2}>$. In this case the scaling function
should be
$\widetilde g(1/2,L/\xi_K(J))+(-1)^{L/2}\widetilde f(1/2,L/\xi_K(J))$,
and we can suppress the first argument. In addition since we only consider
the case where
$L/2$ is even, the scaling relation takes the simpler form
$L<S^z_{L/2}>=h(L/\xi_K)$.
In Fig.~\ref{fig:l2scale} we show $L<S^z_{L/2}>$ for the
coupling constants $J=0.3,0.4,0.5,0.75,1,1.5,1.8,2,2.5,3,3.5,4,10$
as a function of $L/\xi_K$, beginning
with $L=2$. The data can be collapsed onto a single curve
thus determining $\xi_K(J)$ up to
a multiplicative constant. If we fix one of the correlation lengths
the rest of the correlation lengths are fixed by requiring that
the scaling form be obeyed.
An excellent data collapse is obtained.
All the results in this figure are for the one impurity case
in the ground-state subspace $RPH=1, S^z_T=1/2$, with $m=128$ states.
In principle it is possible to obtain the Kondo length scale, $\xi_K$
as a function of $J$
from this scaling plot. However, since $\xi_K$ varies quite
rapidly with $J$ one can essentially only obtain qualitative results
at weak couplings where $\xi_K$ is several thousand lattice spacings or
at strong couplings where $\xi_K$ is very small.
However, for a fair range of intermediate couplings $\xi_K$ can be
extracted with a reasonable precision.
Our results are summarized in Table~\ref{tab:xi}.
We note that Fig.~\ref{fig:l2scale} clearly
displays the complete crossover from weak to strong coupling.

\begin{figure}
\centering
\epsfysize=12 cm
\leavevmode
\epsffile{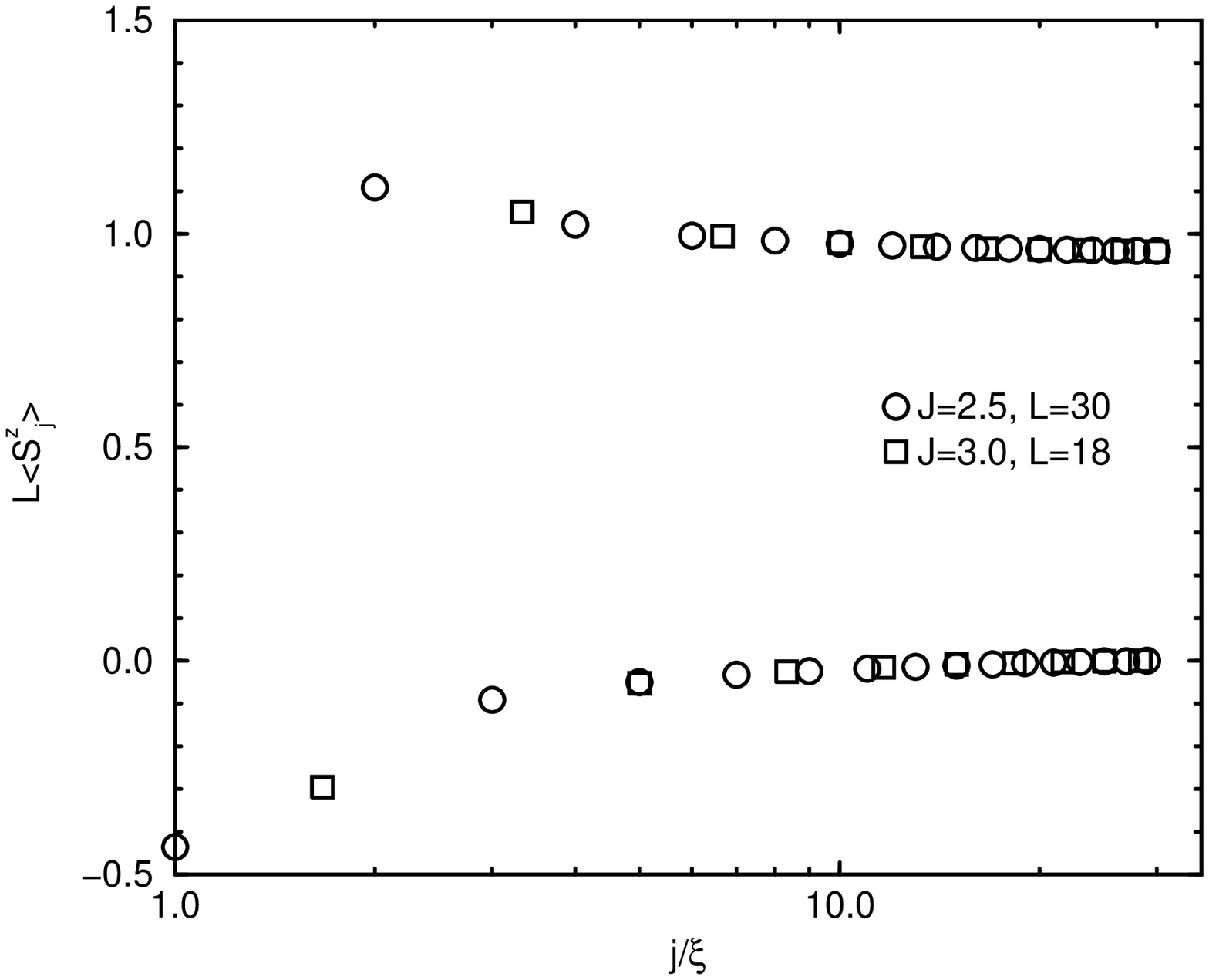}
\caption{$L$ times the expectation value of the z-component of the
electron spin, $<S^z_j>$, as a function of $j/\xi_K(J)$.
Two systems are shown: $J=2.5,\ \xi_K=1.0,\ L=30$ and $J=3.0,
\ \xi_K=0.6,\ L=18$
Thus in both cases we have $L/\xi_K\approx 30$.
Clearly the data collapses onto
a universal curve.
}
\label{fig:lsziscale}
\end{figure}

\begin{figure}
\centering
\epsfysize=12 cm
\leavevmode
\epsffile{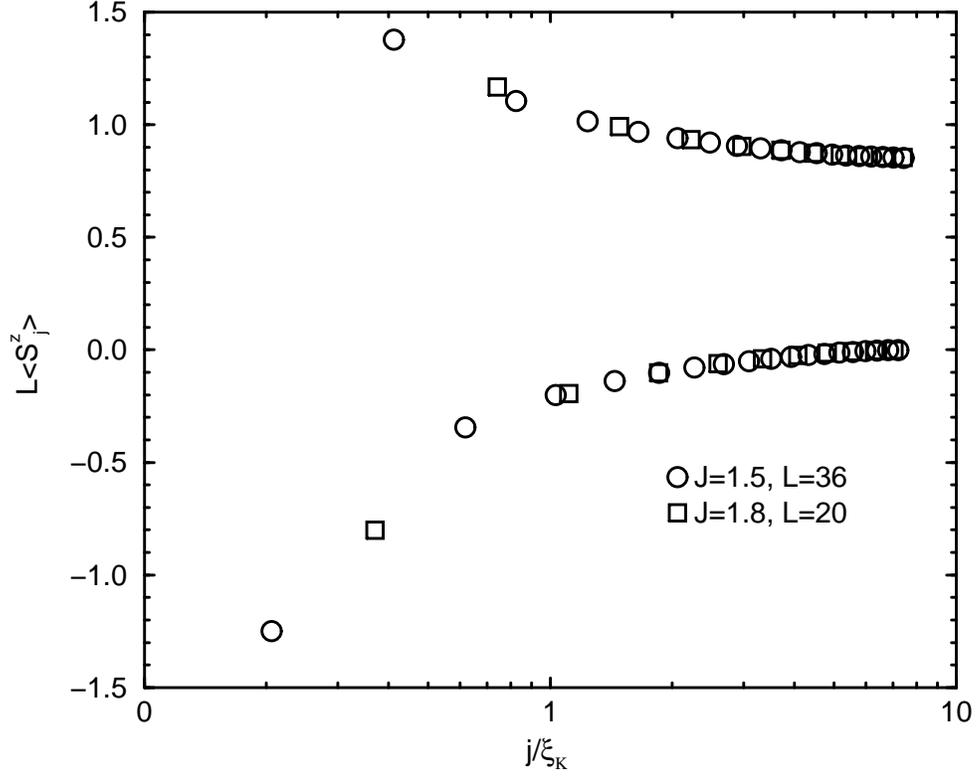}
\caption{$L$ times the expectation value of the z-component of the
electron spin, $<S^z_j>$, as a function of $j/\xi_K(J)$.
Two systems are shown: $J=1.8,\ \xi_K=2.7,\ L=20$ and
$J=1.5,\ \xi_K=4.85,\ L=36$
Thus in both cases we have $L/\xi_K\approx 7.4$.
Clearly the data collapses onto
a universal curve.
}
\label{fig:lsziscale2}
\end{figure}

In the same manner we can look at $L<S^z_j>$,
but instead of fixing $j/L$ we can fix $L/\xi_K(J)$. The
scaling form should then be $h(j/\xi_K(J))$, where we have suppressed the
dependence on $L/\xi_K(J)$. Using
$\xi_K(J=2.5)=1.0,\ \xi_K(J=3)=0.6$, such a data collapse is shown in
{}Fig.~\ref{fig:lsziscale} with $L=30,20$ for $J=2.5,3$, respectively.
Thus, $L/\xi_K\approx 32$ is kept fixed.
Viewing $\xi_K(J=2.5)$ and $\xi_K(J=3)$
as fixed from the previous analysis this plot contains no free parameters.
The collapse is excellent. The same rescaling can also be performed
at other couplings. In Fig.~\ref{fig:lsziscale2} we show results
for $J=1.5$ and $J=1.8$ with $\xi_K(J=1.5)\sim 4.85$
and $\xi_K(J=1.8)\sim 2.7$,
where we use $L=36,20$ for $J=1.5,1.8$ respectively. Again
we find an excellent collapse of the data. The complete scaling
shown in Figs.~\ref{fig:lsziscale} and \ref{fig:lsziscale2}
is a highly non-trivial test of the scaling form Eq.~(\ref{eq:magscaling})
and the fact that the numerical results clearly follow the scaling
form lends strong support to the existence of the Kondo length scale.

\subsection{Scaling of $L<S^z_{\rm imp}>$}
\begin{figure}
\centering
\epsfysize=12 cm
\leavevmode
\epsffile{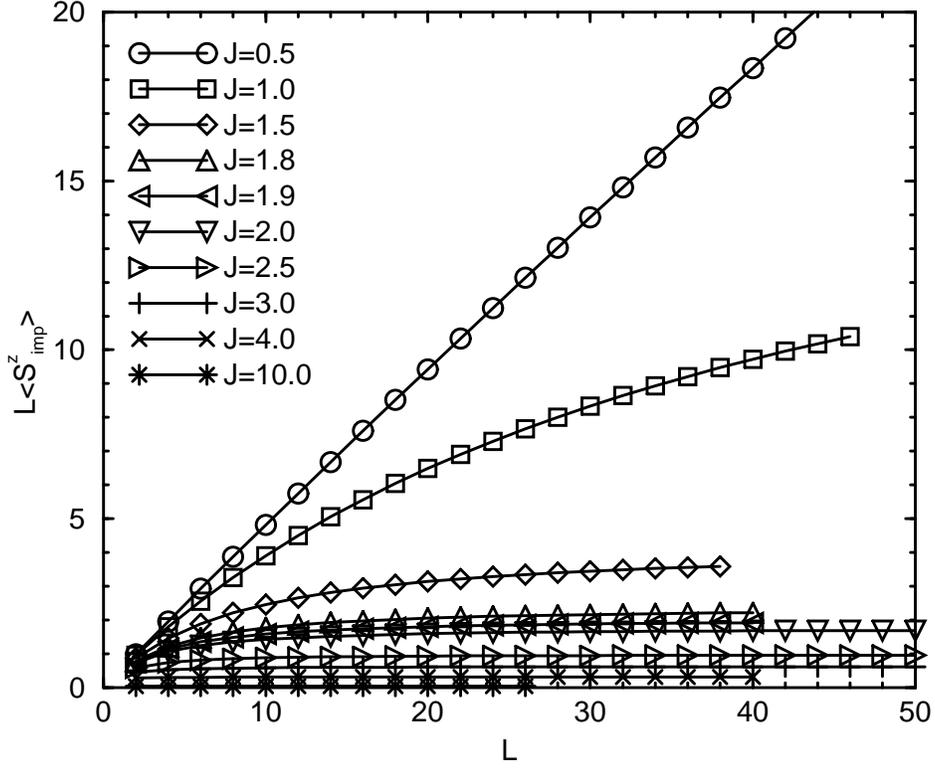}
\caption{The scaled z-component of the impurity spin,
$L<S^z_{\rm imp}>$, as a function
of chain length, $L$, for a range of Kondo couplings $J$. In all
cases does the first point correspond to $L=2$.
}
\label{fig:lsimp}
\end{figure}

We now turn to a discussion of our results for
the expectation value of the impurity spin.
In Fig.~\ref{fig:lsimp} we show $L<S^z_{\rm imp}>$
as a function of $L$ for $J=0.5,1,1.5,1.8,1.9,2,2.5,3,3.5,4,10$,
beginning with $L=2$. As is clearly evident $L<S^z_{\rm imp}>$
approaches a constant in the strong coupling limit. At weak coupling
the behavior is consistent with $L<S^z_{\rm imp}>\sim L/2$ for the
values of $L$ accessible.
{}From the results in Fig.~\ref{fig:lsimp} we can also
check the result Eq.~(\ref{eq:simp}). For $J=10.0$ we find
with the DMRG method that $L<S^z_{\rm imp}> \sim 0.0456$ for
large $L$. This can be compared to $0.0444$ from Eq.~(\ref{eq:simp}),
in very good agreement.

\subsection{$\xi_K$}
In the previous section we obtained numerical results for $\xi_K(J)$
by requiring our numerical results to scale.
We can now try
to fit these results to expressions for $\xi_K$ obtained from
renormalization group arguments which to first order gives
Eq.~(\ref{xi_K}).  Is this possible ?
The first point to realize is that we can only expect the weak coupling
RG formula to
work for a range of J such that $1\ll \xi_K \ll L$.
If $\xi_K$ is too small
then the coupling constant is too big so low order perturbation
theory doesn't work.  If $\xi_K$ is too big then
finite size effects will dominate.

We proceed by obtaining a higher order expression for $\xi_K(J)$
than the simple exponential relation Eq.~(\ref{xi_K}).
To go to one higher order, we first of all need to calculate the
renormalized coupling to $O[(J\rho )^2]$, obtained by reducing the
effective bandwidth (in momentum space) from $\pi /a$ to $2/\xi_0$,
where $\xi_0$, is some length scale much bigger than $a$, the lattice
spacing.  This involves the integral:
\begin{equation}
\int _0^{\pi /2a - 1/\xi_0}{dk\over \cos ka}\propto \ln [\tan
(2\xi_0/a)].
\end{equation}
Thus:
\begin{equation}
\lambda_0 = J\rho + (J\rho )^2\ln [\tan (2\xi_0/a)].
\end{equation}
Here $\rho$ is the density of states,
\begin{equation}
\rho = 1/2\pi t.
\end{equation}
Next we need to integrate the $\beta$-function, to third order.
The $\beta$-function
is~\cite{migdal}:
\begin{equation}
-d\lambda /d\ln \Lambda = \lambda^2-\lambda^3/2.
\end{equation}
We now integrate this equation, using the bare cut-off,
$\Lambda_0=1/\xi_0$, the bare coupling $\lambda_0$ given above, the
renormalized coupling some number c of $O(1)$
and the new cut-off $1/\xi_K$, the
inverse correlation length.  This gives~\cite{arnold}:
\begin{equation}
\xi_K = \xi_0e^{1/\lambda_0-1/c}\sqrt{1-2/\lambda_0\over 1-2/c}.
\label{eq:xij}
\end{equation}
This, together with the equation above for $\lambda_0$ determine
$\xi_K$ vs. J to a better accuracy than the simple exponential
form.  We now have 2 free parameters; c which should be
positive and $O(1)$ and $\xi_0$ which should be $\gg 1$ but $\ll \xi_K$.
This form should be valid for the range of $J$ where
$1 \ll \xi_K \ll L$.
The corrections
arising from the constant $c$ is presumably only one of several equally
important terms. We have included it here to improve the agreement
with the numerical results. In the limit
$\lambda_0 \to 0$ Eq.~(\ref{eq:xij}) reduces to:
$\xi_K \propto e^{1/\lambda_0}/\sqrt{\lambda_0}$,
in agreement with Ref.~\onlinecite{zawa}.

\begin{figure}
\centering
\epsfysize=12 cm
\leavevmode
\epsffile{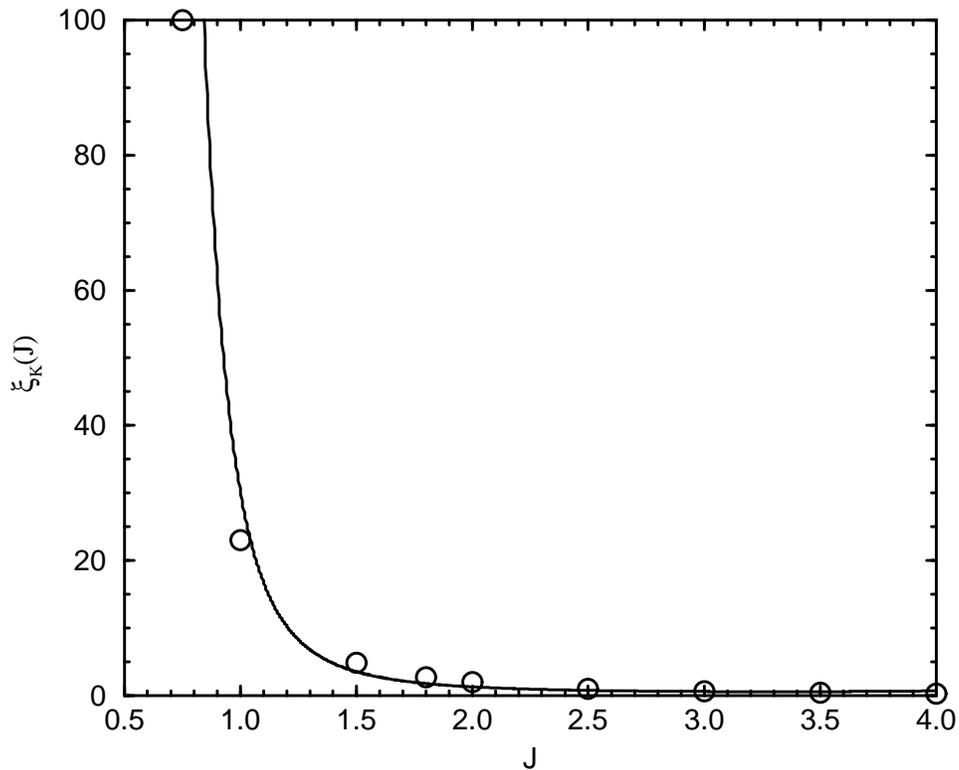}
\caption{The Kondo length, $\xi_K$, as a function of Kondo
coupling, $J$. The circles denote the numerical DMRG results.
The solid line indicates a least square fit of the results shown
to the form Eq.~(\protect\ref{eq:xij}). The fitted parameters
are $\xi_0=1.76,\ c=0.21$.
}
\label{fig:xi}
\end{figure}

A least square fit of the numerical results for $\xi_K(J)$, in
Table~\ref{tab:xi}, to the form Eq.~(\ref{eq:xij}) is shown in
{}Fig.~\ref{fig:xi}. The circles indicate the DMRG results while
the solid line is Eq.~(\ref{eq:xij}) with the fitted parameters
$\xi_0=1.76,\ c=0.21$. Taking into account the sizable uncertainty
in $\xi_K$ at weak coupling the fit works extraordinarily
well.

\section{Conclusion}
We have shown that the local susceptibility can
be brought into a unified scaling form linking a high
temperature RRKY form and a low temperature local Fermi
liquid form. In this picture the Knight shift is longer
range at low temperatures where the screening cloud
has formed, than at high temperatures where it has not.
The experimentally observed behavior of the Knight
shift~\cite{slichter}
can possibly be explained by a factorization of the
scaling functions deep inside the screening cloud which is
the only region the experiments probe.

Non-trivial tests of the scaling form have been performed
by numerically calculating the on-site magnetization
at zero temperature
for finite systems. In all cases scaling behavior
consistent with the proposed form is observed.
Both $<S^z_{L/2}>$ as a function of $L$, and
$<S^z_j>$ as a function of $j$ for fixed $\xi_K/J$
clearly show scaling. Numerical estimates
of $\xi_K(J)$ has been extracted in good agreement with
an estimate of $\xi_K(J)$ from renormalization group
calculations.

\acknowledgments
We gratefully acknowledge many useful discussions with J.~Gan.
We would also like to thank
V.~Barzykin, W.~J.~L.~Buyers, S.~M.~Girvin,
B.~I.~Halperin, A.~H.~MacDonald, N.~Prokof'ev,
P.~C.~E.~Stamp, C.~Varma, E.~Wong and
A.~Zawadowski
for helpful comments. ESS is supported by NSF grant number NSF DMR-9416906.
This research was supported in part by NSERC of Canada.

\begin{table}
\caption{The values for the Kondo length, $\xi_K$, used
in Fig.~\protect\ref{fig:l2scale} and shown
in Fig.~\protect\ref{fig:xi} (circles).}
\label{tab:xi}
\begin{tabular}{dc}
\multicolumn{1}{d}{J}
&\multicolumn{1}{c}{$\xi_K(J)$}\\
\tableline
10.0 &      0.031\\
4.0 &     0.27\\
3.0 &      0.6\\
2.5 &     1.0\\
2.0 &     2.0\\
1.8&	2.7\\
1.5&	4.85\\
1.0&	23\\
0.75&	100\\
0.5&	1200\\
0.4&	7000\\
0.3&	100000\\
\end{tabular}
\end{table}

\eject
\eject

\end{document}